\documentclass[reprint,twocolumn,showpacs,preprintnumbers,amsmath,amssymb,aps,prb,floatfix,longbibliography,superscriptaddress,floatfix,10pt]{revtex4-1}

\usepackage{graphicx}
\usepackage{dcolumn}
\usepackage{bm}
\usepackage{dcolumn}
\usepackage{bm}
\usepackage[usenames]{color}
\usepackage[%
  colorlinks=true,
  urlcolor=blue,
  linkcolor=blue,
  citecolor=blue
]{hyperref}
\usepackage{pifont}
\usepackage[utf8]{inputenc}
 
\begin{document}

\title{Tailoring dual reversal modes by helicity control in ferromagnetic nanotubes}

\author{H. D. Salinas}

\email{hernan.salinas@udea.edu.co}

\affiliation{Grupo de Magnetismo y Simulaci\'on G+, Instituto de
  F\'isica, Universidad de Antioquia. A.A. 1226, Medell\'in, Colombia}

\author{J. Restrepo}

\email{johans.restrepo$@$udea.edu.co}

\affiliation{Grupo de Magnetismo y Simulaci\'on G+, Instituto de
  F\'isica, Universidad de Antioquia. A.A. 1226, Medell\'in, Colombia}

\author{\`Oscar Iglesias} 

\email{oscar@ffn.ub.es}

\homepage{http://www.ffn.ub.es/oscar, http://nanomagn.blogspot.com}

\affiliation{Departament de F\'isica de la Matèria Condensada and Institut de
  Nanoci\`encia i Nanotecnologia, Universitat de Barcelona, Av. Diagonal
  647, 08028 Barcelona, Spain}

\date{\today}

\begin{abstract}
We investigate the effects of the competition between exchange ($J$) and dipolar ($D$) interactions on the magnetization reversal mechanisms of ferromagnetic nanotubes. 
Using first atomistic Monte Carlo (MC) simulations for a model with Heisenberg spins on a cylindrical surface, we compute hysteresis loops for a wide range of the $\gamma=D/J$ parameter, characterizing the reversal behavior in terms of the cylindrical magnetization components and the vorticity parameter along the tube length.
For $\gamma$'s close to the value for which helical (H) states are stable at zero applied field, we show that the hysteresis loops can occur in four different classes that are combinations of two reversal modes with well-differentiated coercivities with probabilities that depend on the tube length and radius.
This variety in the reversal modes is found to be linked to the metastability of the $H$ states during the reversal that induce different paths followed along the energy landscape as the field is changed.
We further demonstrate that reversal by either of the two modes can be induced by tailoring the nanotube initial state so that vortices with equal or contrary chirality are formed at the ends, thus achieving low or high coercive fields at will without changing $\gamma$.
Finally,  the results of additional micromagnetic simulations performed on tubes with similar aspect ratio show that dual switching modes and its tailoring can also be observed in tubes with more microscopic dimensions.
\end{abstract}
\maketitle

\section{Introduction}

Magnetic nanotubes have gained in the last years increasing interest from fundamental and technological standpoints due to their double potential functionality arising from their characteristic inner and outer surfaces \cite{Stano_book2018,Ye_crssm2012}.
In particular, the variety of different magnetic states that can be achieved in these structures have been exploited in sensors, logical devices \cite{Cowburn_science2000}, high-density magnetic memories \cite{Parkin_science2008} and even for magnetic hyperthermia and drug release \cite{Alonso_jap2015,Fernandez-Roldan_apl2018,Chen_acs2012}.  
In particular, vortex ($V$) and helical ($H$) states have been reported to occur \cite{Salinas_jsnm2012,Chen_jmmm2007,Chen_jap2010,Chen_jap2011,Escrig_jmmm2007,Biziere_Nanolett2013,Weber_Nanolett2012,Buchter_prl2013,Yamasaki_prl2003,Salinas_scirep2018} and become relevant for magnetic storage purposes since, being flux-closure configurations, they do not produce stray fields, avoiding the consequent leakage of magnetic flux spreading outward from the tube. 
This fact is important in ensembles where flux-closure configurations can suppress non-desirable interactions between nearby nanomagnets \cite{Proenca_jap2013}. Additionally, the core-free aspect of the tubes as compared to their filled counterparts allows in principle a fast reversal process with a coercivity that can be controlled by changing the shape factor or the aspect ratio. 

All these expectations have been encouraged by the rapid advance in the experimental methods for fabrication of magnetic tubes and decoration techniques, which currently allow to synthesize nanocylinders by means of different routes \cite{Nielsch_jap2005,Han_advmat2009,Sui_apl2004,Xu_jpd2008,Daub_jap2007,Garcia_jpc2018} with a high degree of control on their composition and geometry.  
Moreover, advances in imaging and characterization techniques \cite{Streubel_natcomm2015,Donnelly_nature2017,Zimmermann_nanol2018,Vasyukov_NanoLett2018} allow nowadays to visualize magnetic configurations along different stages of the reversal process 
or even to register hysteresis loops of a single-molecule magnet \cite{Guo_science2018}. 
Therefore, there have been some works reporting the field driven magnetic switching mechanisms on individual nanotubes \cite{Buchter_prl2013} having typical lengths on the range from $0.5$ to some tenths of $\mu$m, radii from $100-500$ $\mu$m and thicknesses from $10-70$ $\mu$m \cite{Wyss_prb2017,Mehlin_prb2018,Vasyukov_NanoLett2018,Zimmermann_nanol2018}.
Given the length scales of real tubes, theoretical understanding on this subject has been gained mainly by using micromagnetic simulations \cite{Fernandez-Roldan_nanom2018}. 
Apart from these, analytical calculations based on continuum approximations \cite{landeros_prb2009, Allende_ejpb2008, Lee_jmmm2007,Chen_jap2010, Chen_jap2011,Chen_mater2018, Li-Ying_pssb2010} have also succeeded in describing the main phenomenology. 
Other theoretical methods \cite{Mi_jmmm2016,Canko_ssc2013} or simulation techniques \cite{Albuquerque_prl2002, Landeros_prb2005,Lehtinen_prl2004,Konstantinova_jmmm2008} have also partially addressed these issues.

Even more appealing is the idea of tubular structures having as magnetic entities not atoms but lattices of nanoparticles \cite{Sotnikov_prb2017,Stankovic_nanosc2019}, clusters or even single-molecule magnets \cite{Bogani_angew2009,Giusti_angew2009} that can be achieved by the controlled deposition on a surface template (e.g. carbon nanotubes \cite{Masotti_ijms2013}). 
The possibility to control the spatial ordering of nanoparticles has also been demonstrated by lattice engineering using DNA shells attached to them \cite{Ji_Small2019}.
Moreover, it has been recently shown that the strength of the exchange interactions can be changed to a certain extent by the application of electrical fields \cite{Fittipaldi_NatMater2019}.
In all these cases, it is expected that the magnetic configurations and reversal behavior is dictated by the competition of exchange and dipolar interactions between the macrospins representing the high-moment magnetic entities, depending also on  the underlying lattice where spins are placed. 

In order to model both kinds of tubular structures, one is certainly forced to revert to simulations at the macrospin (atomistic) level in order to study the reversal mechanisms, with the handicap of having to scale down the length scales to the range of few nanometers due to computational limitations caused by large number of degrees of freedom that might be present in samples of real size.
In this paper, we address the study of the reversal modes of a lattice of spins residing on the surface of a nanotube in the presence of a magnetic field. In order to do so, we will present results of MC simulations of hysteresis loops based on an atomistic model as described above and vary the relative strength of dipolar to exchange interactions to see how they influence the reversal processes along the loops. We discover that for a certain range of $\gamma$'s, dual reversal modes appear as a consequence of the metastability of the magnetic configurations formed during the first stages of the magnetization reversal, and we show a route to tailor these modes that can also be mimicked experimentally.
The rest of the manuscript is organized as follows: In Sec. \ref{Model_Sec}, we present the computational details of the model and of the MC and micromagnetic simulations performed. In Sec. \ref{Results_Sec} we start by analyzing the results of the MC simulated hysteresis loops, show a procedure to tailor the different reversal modes, and end up presenting micromagnetic simulation results that demonstrate that a similar phenomenology can show up in tubes with sizes approaching those studied experimentally. We finish with a discussion about the relevance of these results and their applications and present the main conclusions. 


\section{Model and Computational details}
\label{Model_Sec}

\subsection{Monte Carlo simulation}

Finite single-wall nanotubes with free boundary conditions along the main axis of the tube ($z-$axis) were modeled by rolling planar square lattices along (1,1) direction onto a cylindrical geometry in order to get a zig-zag structure which it has been already demonstrated to exhibit a minimum energy configuration compared to a columnar stacked AA realization \cite{Salinas_scirep2018}. Details about such a construction were already reported \cite{Salinas_jsnm2012}. Tube dimensions are determined by the pairs $(N,N_z)$, being $N$ the number of spins per layer and $N_z$ the number of layers corresponding to height or length. In particular, tubes having the dimensions (8,15) and (8,20) were considered, with $N_z/N$ aspect ratios of 1.875 and 2.5 respectively. Our Hamiltonian reads as follows:
\begin{equation}
  {\cal H}=E_{ex}+E_{dip}+E_{Z},
\end{equation}
where $E_{Z}$ is the Zeeman interaction of the spins with a uniform external field $\vec{H}$ applied along the main axis of the tube, $E_{ex}$ stands for isotropic and short-range exchange coupling between  three-dimensional classical Heisenberg nearest neighbours (n.n) spins $ S_x,S_y,S_z)$ :
\begin{equation}
  E_{ex}=-\sum_{\langle i,j \rangle } J_{ij} \vec{S_i}\cdot\vec{S_j},
\end{equation}
being $J_ {ij} = J$ a positive constant value exchange integral accounting for ferromagnetic interaction, $E_{dip}$ is the long-range magnetic dipolar interaction given by:
\begin{equation}
  E_{dip} = D\sum_{i< j}\left(\frac{
      \vec{S_{i}}\cdot\vec{S_{j}}-3(\vec{S_{i}}\cdot\hat{r}_{ij})(\vec{S_{j}}\cdot\hat{r}_{ij})} {|\vec{r}_{ij}|^3}\right),
\end{equation}
where summation is expanded over all the set of pairs of moments in the lattice taking care of counting once pair interactions, $\vec{r}_{ij}$ is the relative vector between $i$ and $j$ positions, whereas the dipolar coupling parameter is given by:
\begin{equation}
D=\frac{\mu_{0}\mu ^2}{4\pi a^3},
\label{Eq_D}
\end{equation}
in energy units, where $\mu_{0} $ and $\mu$ are the magnetic permeability of vacuum and the magnetic moment per spin respectively, and $a$ is the lattice constant.
\begin{figure}[thb]
	\centering
  \includegraphics[width=0.4\columnwidth]{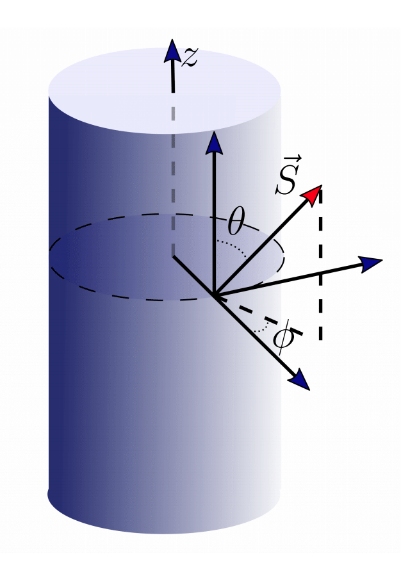}
  \caption{Local reference frame at a spin site where the magnetic moment is given by $\vec{S_i}$ showing the coordinates to describe magnetic configurations, with $\theta$ and $\varphi$ polar and azimuthal angles, repectively.}
\label{Ref-frame_Fig} 
\end{figure}

Thus, we can define the dimensionless parameter $\gamma=D/J$ to quantify the degree of competition between short-range and long-range interaction. 
In order to get an estimate of its possible values for realistic systems, let us first consider spins associated to atomic magnetic moments, 
and let us set $J=10$ meV and the atomic separation $a=0.5$\r{A} . Then, values of $\gamma$ in the range $[0.01, 0.07]$ used in our simulations are obtained for $\mu\approx (1.52-4.03) \mu_B$ which are a typical values for magnetic ions.
On the other hand, the strength of $J$ in nanosystems may be reduced due to the smaller coordination number of atoms at surfaces or interfaces, increasing the relative importance of dipolar interactions. For instance, DFT calculations \cite{Jensen1991} have estimated that the strongest exchange energy per atom in Fe/Ir(111) is less than $5$ meV. Considering a realistic value of $D=0.1$ meV for Fe atoms, this would give a $\gamma$ of the order of $0.02$, which is also in the range of values considered in this work.   

%

Based on the previous Hamiltonian, we have conducted standard Metropolis MC simulations to obtain the hysteresis loops at a fixed temperature $T= 0.1 J/k_B$ by cycling the magnetic field in constant steps $\delta H$ between $0.02$ and $0.1$ depending on the field region, while performing thermodynamic averages of the several components of the magnetization and its rotational as described below. Since thermodynamic states in the hysteresis loops are not strictly equilibrium states, up to $100$ different runs per loop were performed in order to carry out configurational averages with the corresponding error bar calculations. The maximum number of MC steps ($5\times 10^{3}$) and those discarded for thermalization ($3\times 10^{3}$) were kept fixed for all the hysteresis loops.

As already proposed in our previous work \cite{Salinas_scirep2018}, in order to analyze the onset of non-collinear spin configurations during reversal modes, more concretely $V$ or $H$ states, we will use the vorticity order parameter defined as a discretized version of the magnetization curl $\vec{\rho}=\vec{\nabla}\times \vec{M}$, that quantifies the vorticity or degree of circularity of the magnetic configurations (similarly to toroidal moment \cite{Lehmann_NatureNano2019}). Their chirality can be characterized in terms of the sign of the azimuthal component of the magnetization $m_\phi$.
 
We will track the intermediate configurations attained during the reversal process of the hysteresis loops by expressing magnetic moments in usual cylindrical coordinates \cite{Salinas_scirep2018}, that are better suited to describe $V$ and $H$ states. Thus, at every spin site $\vec{r_i}=(R\cos\Phi_i,R\sin\Phi_i,z_i)$ where $R$ is the tube radius and $\phi_i$ is the respective azimuthal coordinate, a local reference frame was considered so the spin vector components read $\vec{S_i}=(S_{z_i},S_{{\phi}_i},S_{{\rho}_i}) =(\cos\theta_i,\sin\theta_i\sin\phi_i,\sin\theta_i\cos\phi_i)$ (see Fig. \ref{Ref-frame_Fig} for angle definitions). Therefore, a full characterization of spin configurations during the switching modes can be obtained both by plotting the average polar angle $\left<\theta\right>$ and the $m_{\phi}$ component per layer for each $z$ value along the entire length of the tube.

\subsection{Micromagnetic simulations}
\label{Micromag_Sec}
In order to investigate the hysteretic behavior of tubes with greater dimensions, we consider a continuous model of the magnetization for analyzing the reversal modes in having an internal diameter of $15$ nm, an external one of  $21$ nm, and a height of $52.5$ nm, where the corresponding aspect ratio is the same as the (8,20) tube. To do so, the Landau-Lifshitz-Gilbert (LLG)  equation was solved by using the Object Oriented Micromagnetic Framework (OOMMF) \cite{OOMMF} where magnetostatic, exchange, anisotropy and Zeeman energies were considered. In particular, simulations were performed for FeCo by using the corresponding parameters, namely, a stiffness constant $A=1.08\times 10^{-11}$ J/m, a saturation magnetization $M_s=1.83\times 10^{6}$  A/m, and we have assumed that the tube has uniaxial magnetic anisotropy in the direction [110] parallel to the symmetry axis $K=0.1\times 10^{5}$J/m$^3$\cite{Bran_jap2013}. Smallest linear size of the cubic cell for space partitioning was set at $1.5$ nm.
\begin{figure}[thb]
 \centering
 \includegraphics[width=0.8\columnwidth]{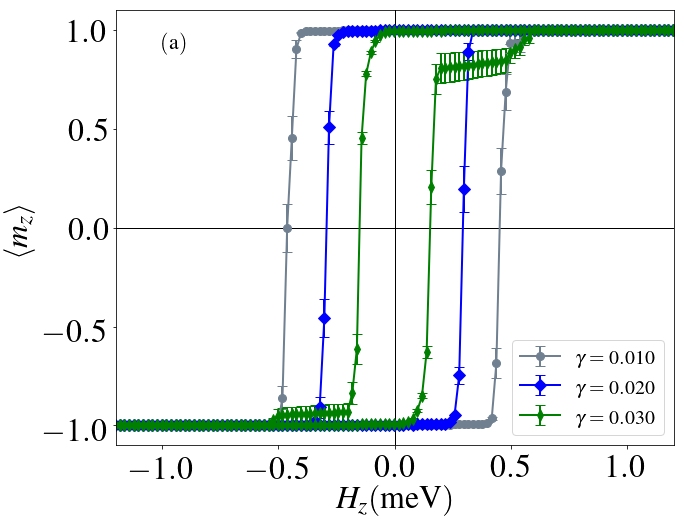}
 \includegraphics[width=0.8\columnwidth]{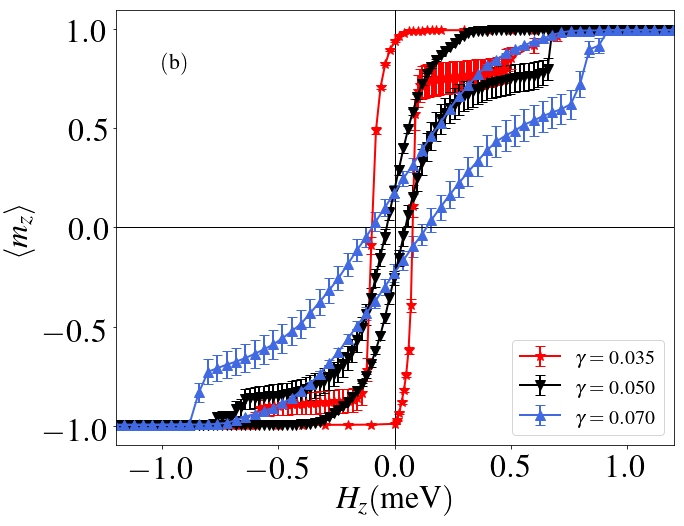}
\caption{ Low-temperature hysteresis loops averaged over 30 different runs for $6$ different values of $\gamma$ for a $(8,15)$ tube  when the field is applied parallel along its axis. }
 \label{Hyst_loopsgammas_Fig}
\end{figure}
\section{Results}
\label{Results_Sec}
\subsection{Hysteresis loops from MC simulations}
In our recent work \cite{Salinas_scirep2018}, we established phase diagrams for the possible equilibrium configurations at zero field for nanotubes with competing exchange and dipolar interactions depending on the value of $\gamma$ and geometric characteristics of the tubes. In particular, independently of the tube length and radius, at low $\gamma$, we found ferromagnetic  (FM) ground states states along the tube axis while, for large enough $\gamma$, $V$ states were found to become stable. Interestingly, for a range of $\gamma$ around a critical value $\gamma^{\star}$ that depends on the geometric parameters of the tubes, states with helical order appear.

Therefore, it is expected that the magnetization reversal mechanisms of the nanotubes under the application of a magnetic field will depend on the value $\gamma$. In order to study this, we will first simulate low-temperature ($T=0.1 J/k_B$) hysteresis loops for several values of $\gamma$ taking as a reference a (8,15) tube, for which we found $\gamma^{\star} \simeq 0.035$ \cite{Salinas_scirep2018}. The results of calculated hysteresis loops averaged over $30$ runs, in which the initial random number generator seed was changed, are shown in Fig. \ref{Hyst_loopsgammas_Fig} for selected values of $\gamma$.
\begin{figure}[thb]
 \centering
 \includegraphics[width=0.8\columnwidth]{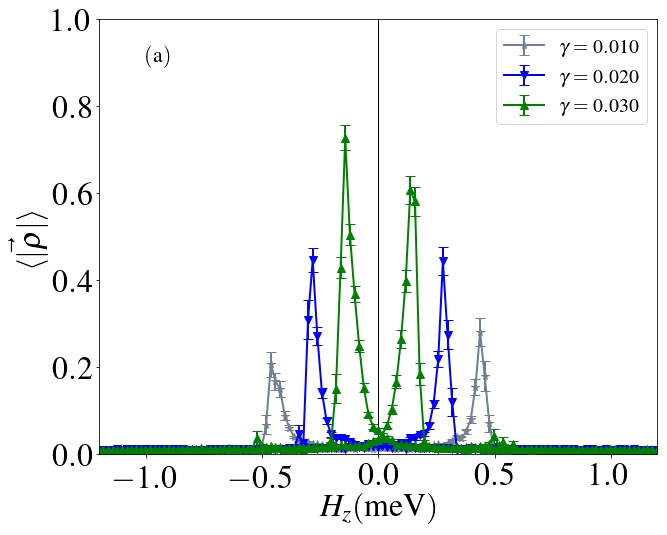}
 \includegraphics[width=0.8\columnwidth]{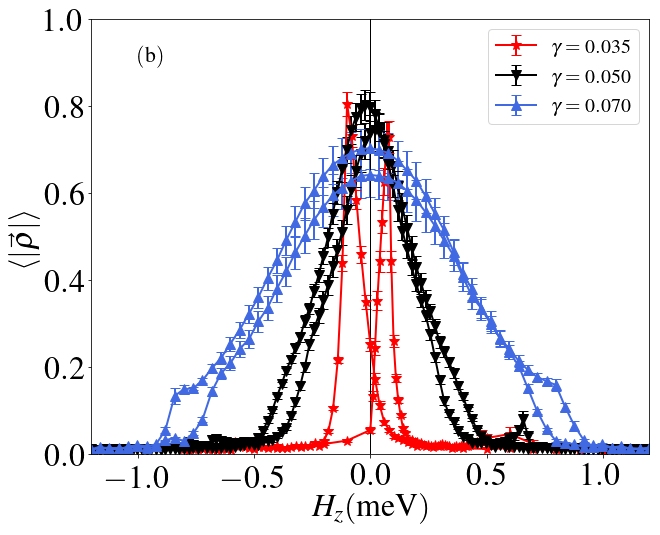}
\caption{ Same as Fig.\ref{Hyst_loopsgammas_Fig} but for the z component of the rotational of the magnetization ${|\rho|}$ .}
 \label{Vort_loopsgammas_Fig}
\end{figure}

More concretely,  for low $\gamma$ values, when the exchange interaction is dominant, the tubes behave as  collinear ferromagnets that reverse their magnetization coherently with hysteresis loops that exhibit a high degree of squareness and appreciable but low values of the vorticity only around the coercive fields. On increasing $\gamma$, the coercive field diminishes as a consequence of the spin canting induced by the dipolar interaction at the tube ends, which facilitates the spin reversal. 
Even though the shape of the hysteresis loops in the range $\gamma < 0.03$ is qualitatively similar, suggesting that the reversal mechanisms are preserved, microscopically some differences progressively emerge as indicated by the rounding of the loops near the coercive fields. 
This can be better appreciated by looking at the variation of the vorticity along the loops shown in Fig. \ref{Vort_loopsgammas_Fig}b, where we observe a progressive increase of the vorticity maxima together with an increase of the width of the peaks around the coercive fields. This indicates the appearance of intermediate circular magnetic states during the switching process, although this is limited to fields close to the coercive field.
For values of $\gamma\approx \gamma^{\star}$, the averaged loops become asymmetric and plateau regions with considerable error bars appear progressively, pointing to variations in the inversion modes form run to run. This fact can be related to the metastability of the $H$ states that can be formed at intermediate states of the reversal process for this range of $\gamma$'s, as mentioned before.

\begin{figure}[thb]
\centering
\includegraphics[width=0.9\columnwidth]{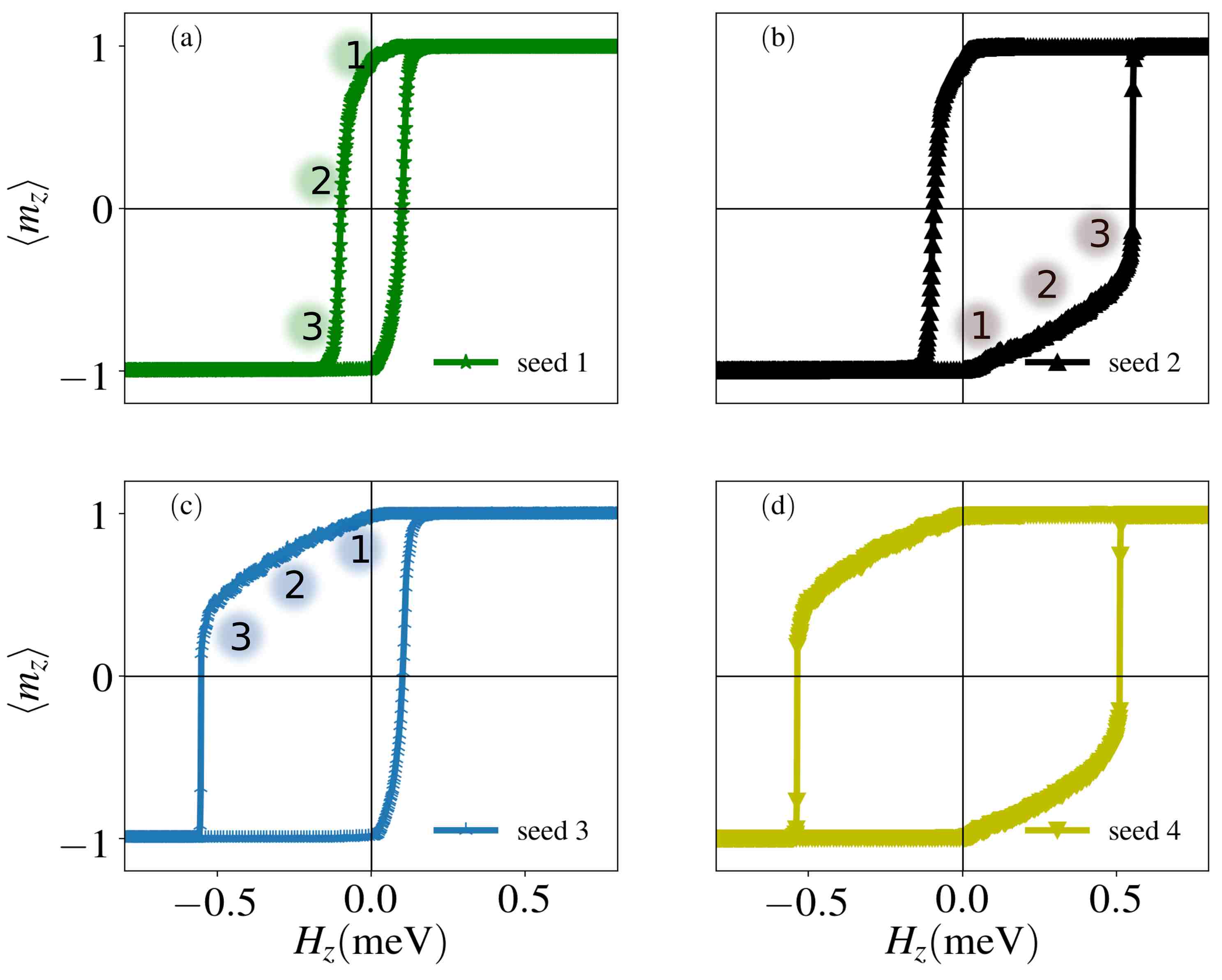}
\caption{The $4$ possible paths followed when a $(8,15)$ nanotube with $\gamma=0.035$ is submitted to an hysteresis loop simulated starting from different seeds of the random number generator. The $4$ cases are named according to the reversal modes followed along the decreasing-increasing field branches: (a) $Q1-Q1$, (b) $Q1-Q2$, (c) $Q2-Q1$, (d) $Q2-Q2$. 
}
\label{Four-paths_8-15_Fig}
\end{figure}
In order to elucidate the origin of these features, $100$ additional different runs were performed for $\gamma=0.035$ by changing the initial seed each time. Results shown in Fig. \ref{Four-paths_8-15_Fig} reveal that all the hysteresis loops without exception, and under the same simulation conditions, fall into four well-defined categories or paths with different probabilities of occurrence. 
A more detailed analysis allows to identify the occurrence of two different switching modes along the decreasing or increasing field branches, namely, one with lower coercivity ($Q1$) and another with higher coercivity ($Q2$). Thus, the four classes of hysteresis loops can be categorized by combinations of them. Two classes ($Q1-Q1$, $Q2-Q2$) correspond to symmetric loops [see panels (a) and (d) in Fig. \ref{Four-paths_8-15_Fig}], while for the other two ($Q1-Q2$, $Q2-Q1$), the cycles are asymmetric, resembling those found in exchange-biased systems.   
Their respective probabilities of occurrence are: $86$\% ($Q1-Q1$), $11$\% ($Q1-Q2$), $2$\% ($Q2-Q1$), $1$\% ($Q2-Q2$); while the total probabilities per mode (branch) are: $92.5$\% for $Q1$ and $7.5$\% for $Q2$.

\begin{figure}[thb]
\centering
\includegraphics[width = 1.\columnwidth]{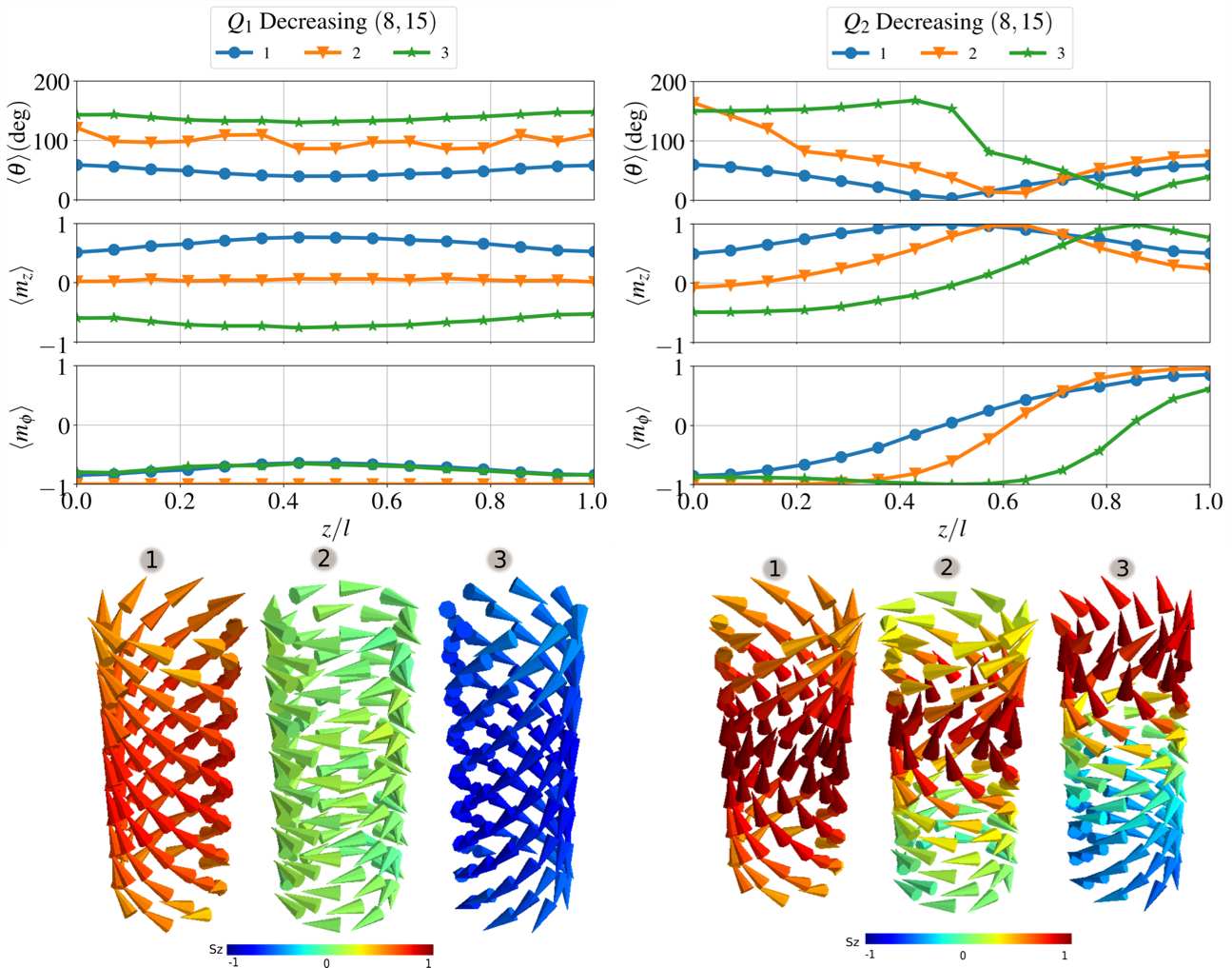}
\caption{Magnetic configurations along the hysteresis loops for the different reversal modes displayed in Fig. \ref{Four-paths_8-15_Fig} (left and right columns correspond to panels (a) and (c) of that figure). Upper panels represent the height profiles of the quantities $\left<\theta\right>$, $\left<m_z\right>$ and $\left<m_{\phi}\right>$  averaged per layer for the tube (8,15) and $\gamma=0.035$, whereas lower ones present snapshots of the spin configurations taken at points labeled in Fig. \ref{Four-paths_8-15_Fig}.}
\label{Confs_8-15_Fig}
\end{figure}

In the $Q1$ mode, the reversal is initiated with the formation of an $H$ state in which both tube ends have the same chirality (see snapshot $1$ in Fig. \ref{Confs_8-15_Fig}, left column). Due to this, the magnetization switching  proceeds in a completely coherent fashion though a gradual rotation of the $H$ angle $\theta$ at both tube ends, reaching a state at remanence (snapshot $2$) that  corresponds to an almost perfect vortex. As can be seen in 
Fig. S1
of the Supplemental Material \cite{Suppl}, the broad peak of $\left<\rho_z\right>$ near the coercive fields and its value close to $1$ corroborates this. The formation of this $H$ and gradual transition into a vortex is responsible for the low associated coercive field. This behavior is confirmed by the profiles shown in the upper left panels of Fig. \ref{Confs_8-15_Fig}, where for stage $2$  $\left<\theta\right>\approx 90^{\circ}$, $\left<m_z\right>\approx 0$ and $\left<m_{\phi}\right>\approx -1$ values are attained. 
At difference from this, in the $Q2$ mode, the reversal is started by the formation of an $H$ state having opposite chiralities at the tube ends  as can be observed in snapshot 1 in Fig. \ref{Confs_8-15_Fig} (central column). As the reversal progresses, $H$ states propagate as $V$ states of opposite chirality by forming a vortex-antivortex pair (stage 2 in the right column panels of Fig. \ref{Confs_8-15_Fig}), which is confirmed by the positive and negative values of the $m_{\phi}$ component close to $1$ and $-1$ respectively, as shown in the respective profiles of right panels of Fig. \ref{Confs_8-15_Fig} . The formation of a non-collinear or $H$ region in the middle part of the tube due to the confrontation of the opposite chiralities makes the system magnetically harder and, therefore, a higher coercive field is obtained. Notice also that (as can be seen in Fig. S1 
of the Supplemental Material \cite{Suppl})  the peak of $\left<\rho_z\right>$ near the coercive fields is narrower and clerly less than $1$ at difference with what happens for the $Q1$ mode.

In Figs. S2-S5 of the Supplemental Material \cite{Suppl}, we provide details about the different reversal modes and configurations for $(8,15)$ tubes with other value of $\gamma$, corresponding to quasi-uniform reversal ($\gamma=0.01, 0.03$) and reversal through V states ($\gamma=0.05, 0.07$).

The occurrence of different paths for reversal can be enlightened by studying the variation of the exchange and dipolar energies of the configurations attained during the decreasing field branch of the hysteresis loops of Fig. \ref{Four-paths_8-15_Fig} (a) and (c), which are shown in Fig. \ref{Energies_MC_Fig} as a function of the corresponding magnetization (see Fig. S6 
 in the Supplemental Material \cite{Suppl} for the same curves as a funtion of the magnetic field). 
For path $Q1$, while going from saturation to the coercive field, the dipolar energy is continuously and progressively decreased at the expense of an increase in exchange energy, explaining the formation of a $V$ state at $H_c \approx 0.1$ meV. 
However, path $Q2$ is characterized by a an excursion through intermediate states with higher total energies and this explains the lower probabilties of occurence compared to path $Q1$. 
Although, in both cases, reversal starts with identical decrease (increase) of the dipolar (exchange) energy, differences between paths are initiated near remanence, when the $Q2$ path drives the system through a path passing a state with $\left< m_z\right>\approx 0.5$ (see snapshot $2$ with $H$ order in the right column of Fig. \ref{Confs_8-15_Fig}) which has a local minimum in $E_{dip}$ that is higher than for the $Q1$ path. 
Near $\left< m_z\right>= 0$, the $Q2$ path displays an abrupt decrease in both $\varepsilon_{ex}$ and $\varepsilon_{dip}$ which corresponds to the two vortices with opposite chirality passing each other at the center of the tube.

Finally, it is worth mentioning that when dipolar interactions dominate over exchange ($\gamma> \gamma^{\star}$),  as can be seen in Fig.\ref{Hyst_loopsgammas_Fig}(b), the averaged loops become more tilted, with higher closure fields and with coercivity tending to zero. This is a consequence of the formation of almost perfect $V$ states at remanence, which for this range of $\gamma$ are the minimum energy configurations \cite{Salinas_scirep2018}, as can be seen in Figs. S4 and S5 
of the Supplemental Material \cite{Suppl}. Two possible reversal paths are also found in this case, again depending on the formation of a state with equal or opposite chiralities at the early stages of reversal. However, now one of the paths has zero coercive field and corresponds to the formation of a state with progressively increasing vorticity with a maximum close to $1$ in zero field (see the $\gamma= 0.05, 0.07$ cases in Figs. \ref{Hyst_loopsgammas_Fig} and \ref{Vort_loopsgammas_Fig}).
\begin{figure}[tbh]
\centering
\includegraphics[width=0.7\columnwidth]{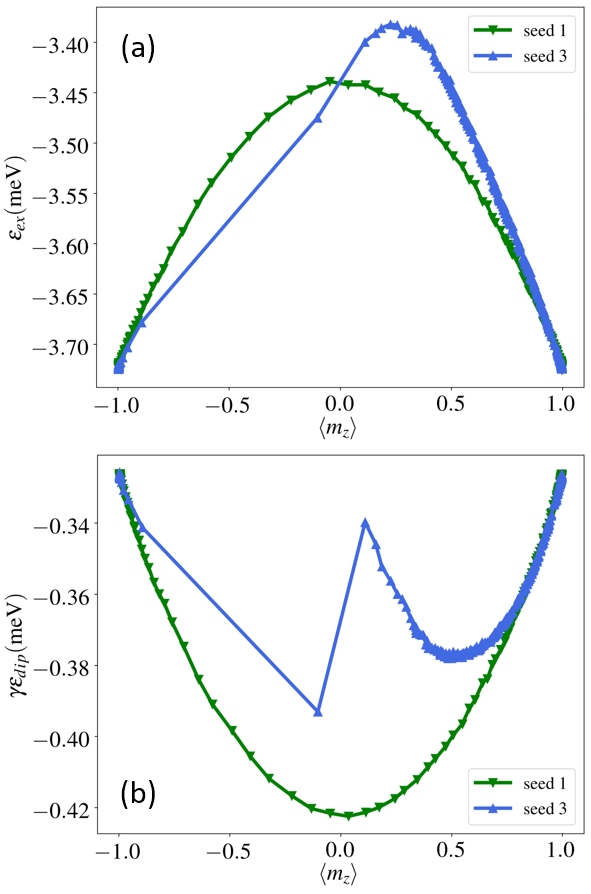}
\caption{Dependence of the exchange (a) and dipolar (b) energies of the (8,15) nanotube on its magnetization for the configurations attained along the decreasing field branch of the hysteresis loops shown in Fig. \ref{Four-paths_8-15_Fig} (a) and (c) (green and blue symbols, respectively).}
\label{Energies_MC_Fig}
\end{figure}

\subsection{Tailoring the reversal modes}

In order to verify the reproducibility of the paths observed in the $Q1$ and $Q2$ modes, we examined the possibility to induce the reversal of the nanotubes by either of the two modes by preparing them in two initial configurations before the field reversal, both consisting in a central region where spins are aligned along the tube length  ($\left<\theta\right>= 0^{\circ}$, $\left<m_z\right>= 1$ and $\left<m_{\phi}\right>= 0$) and $H$ order in the last two layers near the tube ends. 
In one case, the two ends have the same chirality  [$\left<\theta\right>= 45^{\circ}$, $\left<m_z\right>= \cos(45^{\circ})$ and $\left<m_{\phi}\right>= -\sin(45^{\circ})$] , i.e. with the same sign of $\left<m_{\phi}\right>$ at the ends, and in the other they have opposite chirality ($\left<\theta\right>= 45^{\circ}$, $\left<m_z\right>= \cos(45^{\circ})$ and $\left<m_{\phi}\right>= \pm\sin(45^{\circ})$). Due to the cylindrical symmetry, both states result in the same total magnetization [$\left<|\vec{m}|\right>=\left<m_z\right>= 0.92$, $\left<m_x\right>=\left<m_y\right>=0$] and energy ($\left<E\right>=-4.5$ meV) per spin. 

\begin{figure}[thb]
\centering
\includegraphics[width=0.8\columnwidth]{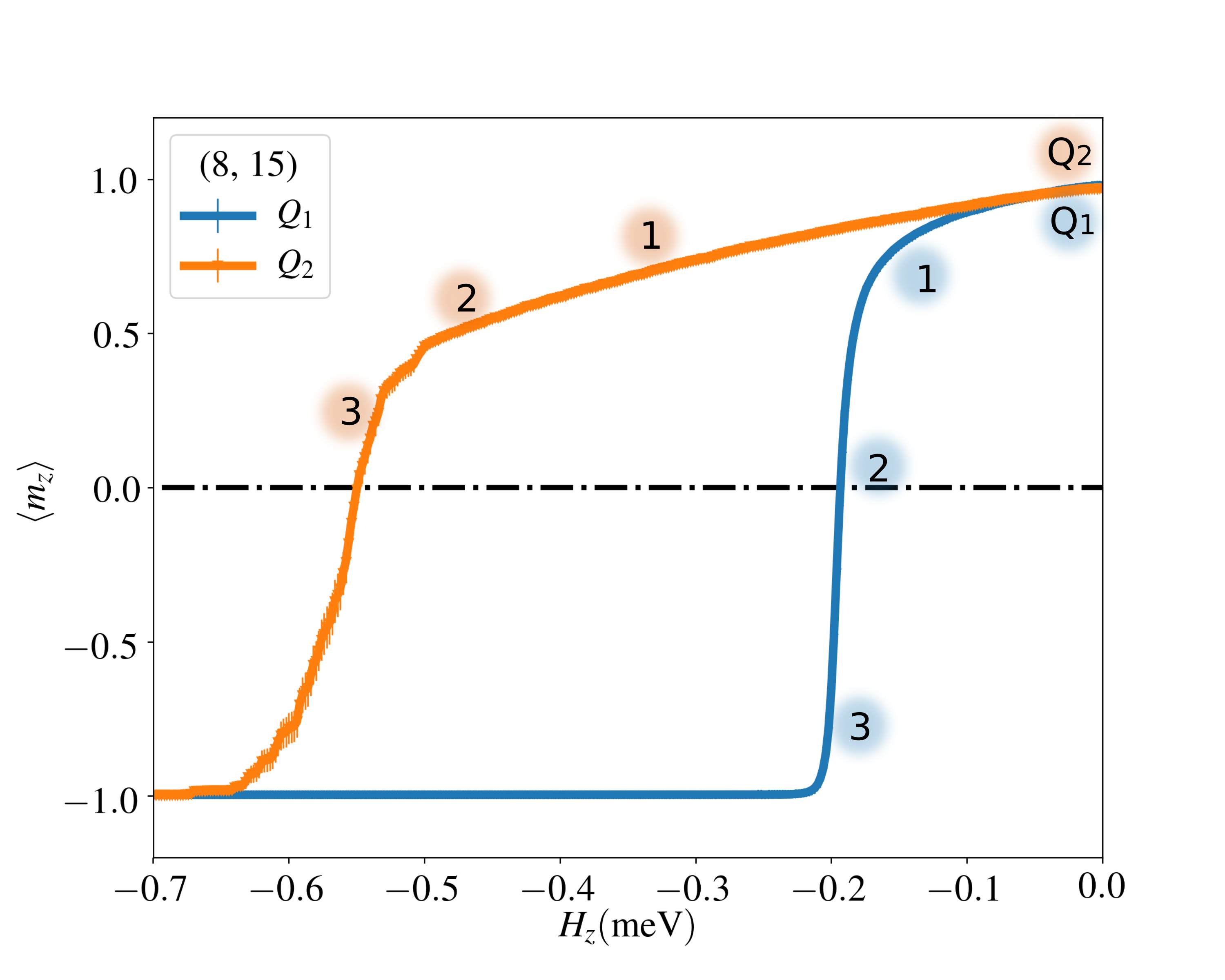}
\caption{Decreasing field branch of the hysteresis loops for a (8,15) tube with $\gamma=0.035$ for two states initially prepared with $H$ states of the same ($Q1$) or opposite  ($Q2$) chiralities at the tube ends.}
\label{Q1-Q2_8-15_Fig}
\end{figure}

\begin{figure}[tbh]
\centering
\includegraphics[width=1.\columnwidth]{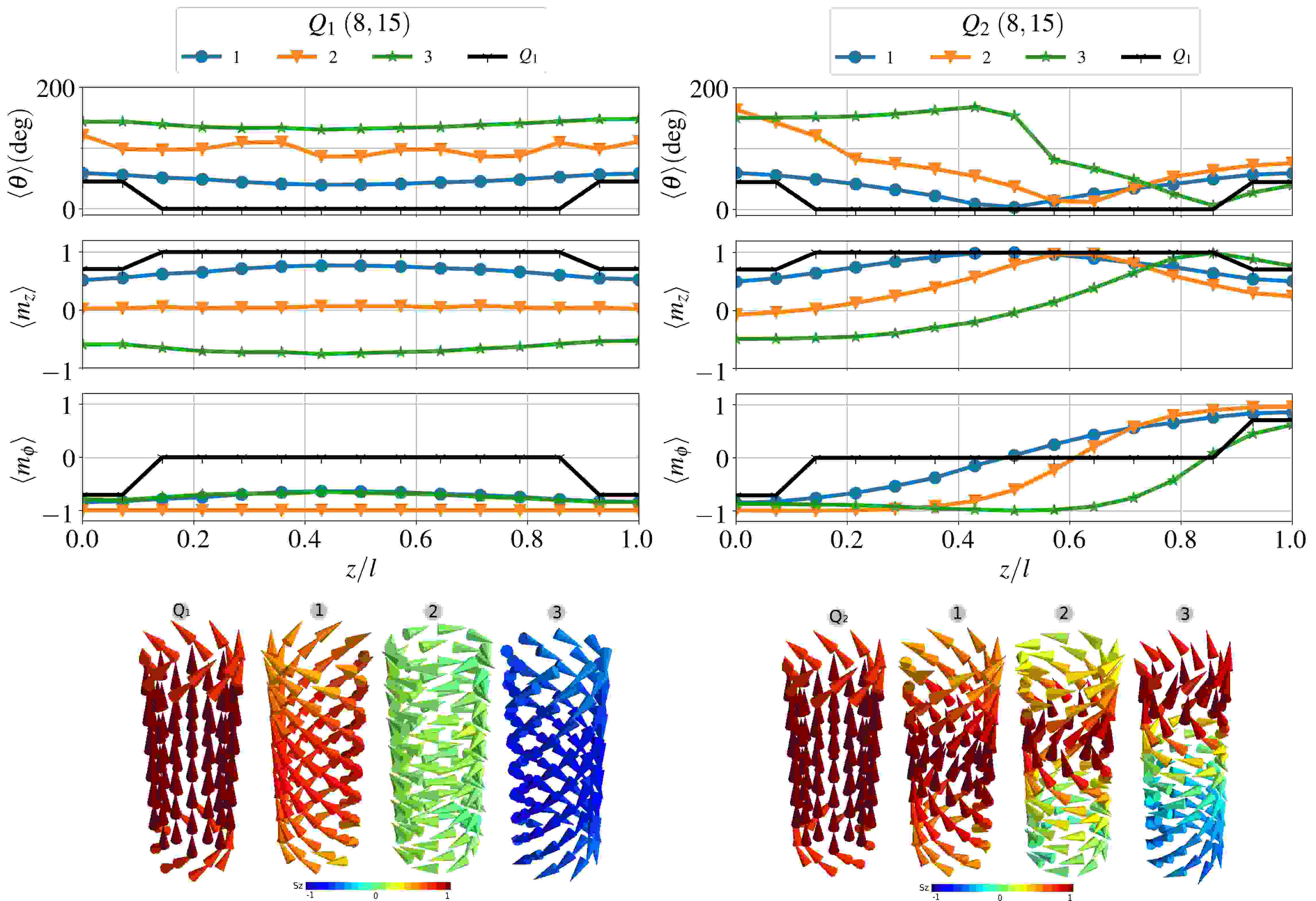}
\caption{Magnetic configurations during magnetization reversal when starting from an initial state $Q1$ ($Q2$) prepared with the same (opposite) chirality at the ends of the tube.
Upper panels represent the height profiles of the quantities $\left<\theta\right>$, $\left<m_z\right>$ and $\left<m_{\phi}\right>$  averaged per layer for the tube (8,15) and $\gamma=0.035$, whereas lower ones present snapshots of the spin configurations taken at the points labeled in Fig.  \ref{Q1-Q2_8-15_Fig}. Profiles of the initial remanence prepared state are included for comparison.
}
\label{Q1-Q2_confs8-15_Fig}
\end{figure}

These initial states were then allowed to evolve under the same conditions along the hysteresis loops with decreasing field. Results are summarized in Fig. \ref{Q1-Q2_8-15_Fig} and the respective initial prepared states along with some intermediate spin configurations during magnetization reversal are shown in Fig. \ref{Q1-Q2_confs8-15_Fig}. 
Starting from the state with the same chirality at both ends, the system always evolves through the $Q1$ inversion mode of lower coercivity, i.e. $P(Q1)=100$\%. However, starting from the configuration with opposite chiralities, the probability to follow the $Q2$ mode is only $P(Q2)=15$\% and there is a $P(Q1)=85$\% probability of observing reversal  through the $Q1$ mode.
These features reveal the subtleties and complexity of the energy landscape and the differences in energy between $Q1$ and $Q2$ modes.  

\begin{figure}[htb]
\centering
\includegraphics[width=1.0\columnwidth]{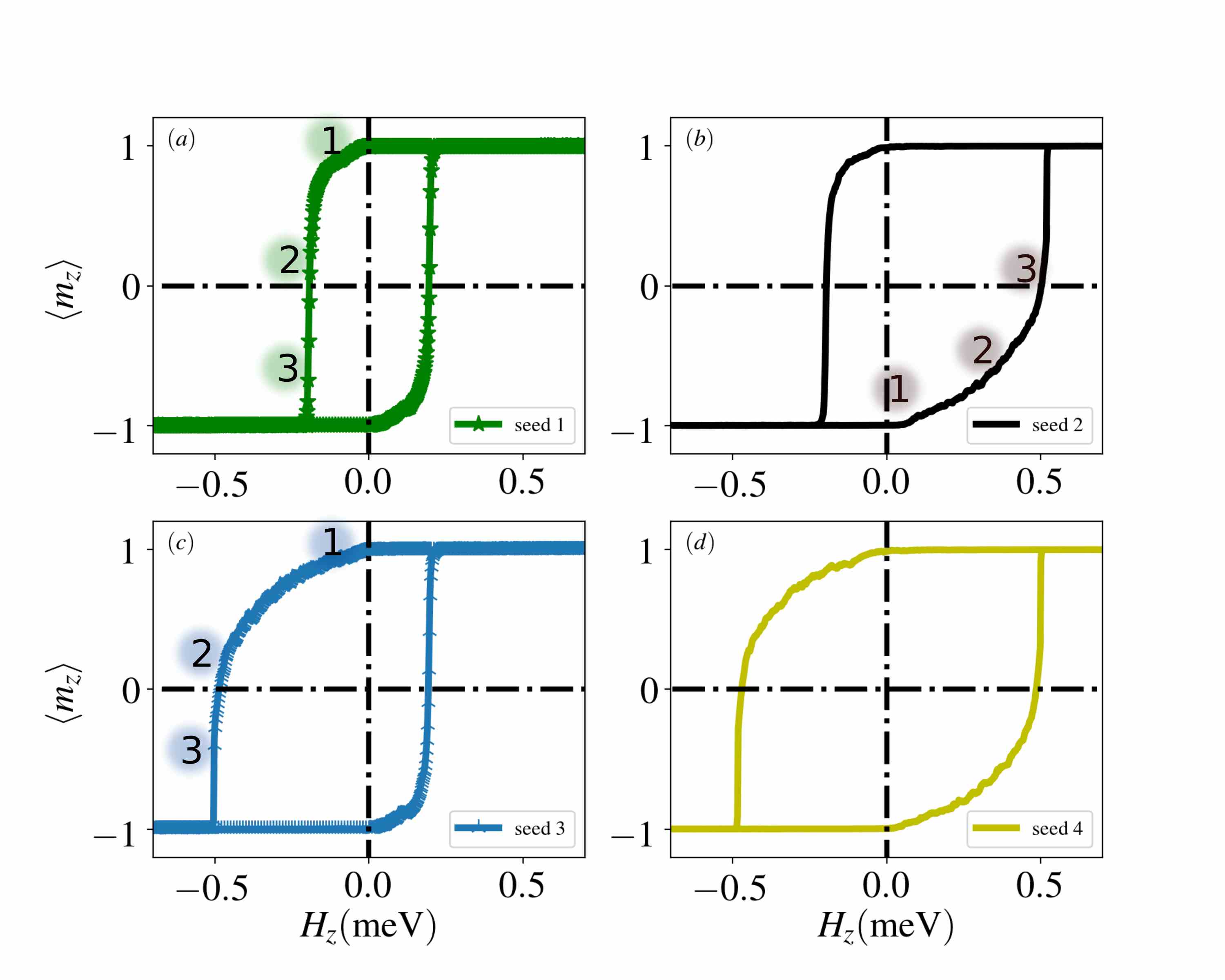}
\caption{The $4$ possible paths followed when a $(8,20)$ nanotube with $\gamma= 0.035$ is submitted to an hysteresis loop simulated starting from different seeds of the random number generator. The $4$ cases are named according to the reversal modes followed along the decreasing-increasing field branches: (a) $Q1-Q1$, (b) $Q1-Q2$, (c) $Q2-Q1$, (d) $Q2-Q2$.} 
\label{Four-paths_8-20_Fig}
\end{figure}

\begin{figure}[htb]
\centering
\includegraphics[width = 1.\columnwidth]{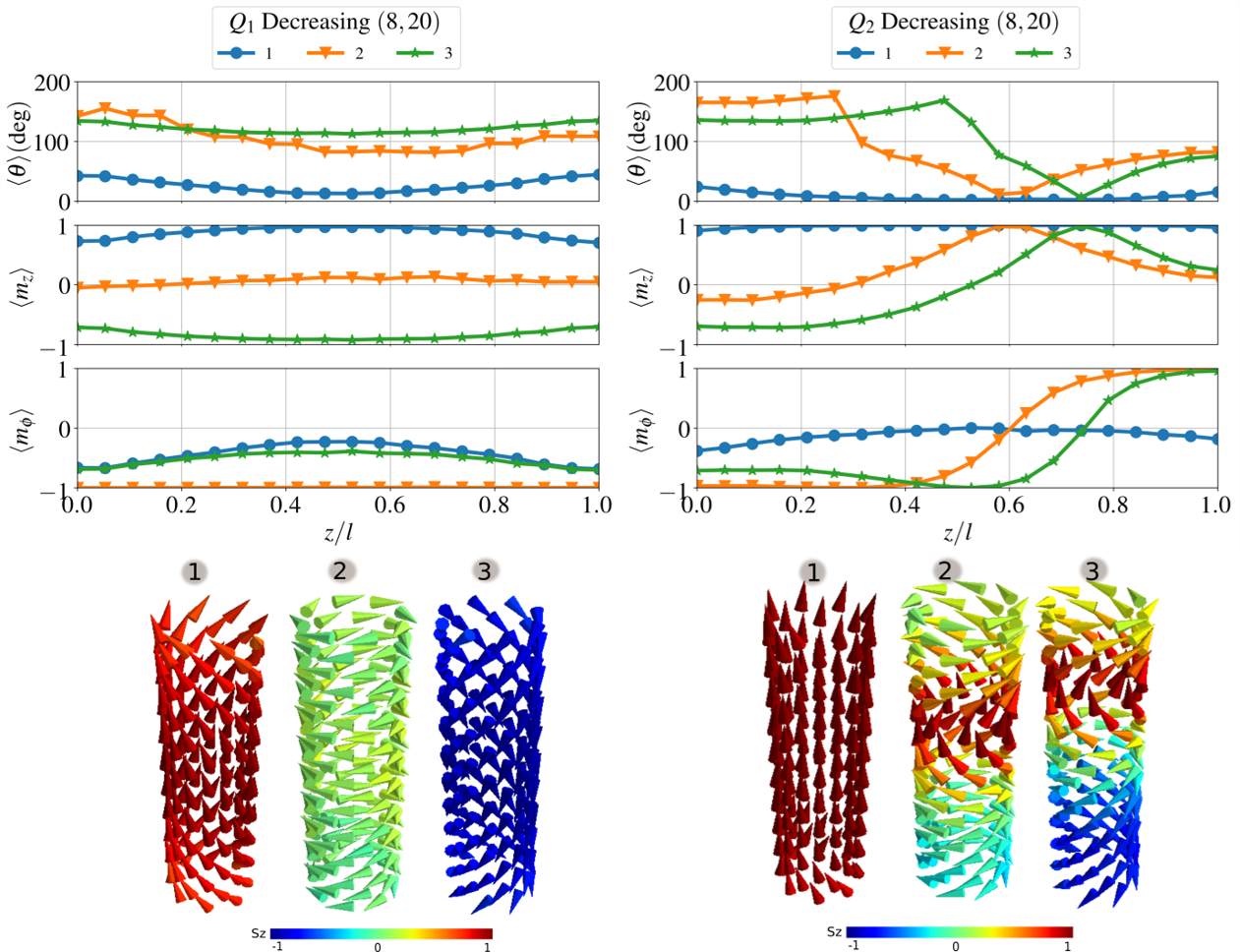}
\caption{Magnetic configurations along the hysteresis loops for the different reversal modes displayed in Fig. \ref{Four-paths_8-20_Fig} (left and right columns correspond to panels (a) and (c) of the figure). Upper panels represent the height profiles of the quantities $\left<\theta\right>$, $\left<m_z\right>$ and $\left<m_{\phi}\right>$  averaged per layer for the tube (8,20) and $\gamma=0.035$, whereas lower ones present snapshots of the spin configurations taken atpoints labeled in Fig. \ref{Four-paths_8-20_Fig}.}
\label{Confs_8-20_Fig}
\end{figure}

Based on the previous results, we have investigated the effect that the tube length may play in the observed metastability of the reversal modes. For this purpose, additional simulations were performed for a $(8,20)$ tube. The results evidence the same phenomenology as for the (8,15) tube (see Fig. \ref{Four-paths_8-20_Fig} and \ref{Confs_8-20_Fig}), although now the coercive field of the $Q1$ mode has increased whereas that of the $Q2$ mode has decreased compared to the $(8,15)$ tube. Moreover, differences can be observed in the probabilities of occurrence of the $Q1$ and $Q2$ modes, which are now $31\% (Q1-Q1)$, $26 \% (Q1-Q2)$, $26 \% (Q2-Q1)$, $17 \% (Q2-Q2)$ with total probabilities per mode  $P(Q1)=57 \%$ and $P(Q2)=43\%$. Notice also that the four reversal paths have become close to equiprobable, which contrasts with the results for the $(8,15)$ tube, where the $Q1-Q1$ path was predominant.
More importantly, when repeating the simulations with differently set initial states, we now obtain $100 \%$ reversal through the $Q1$ and $Q2$ modes when starting from a state with the same or opposite chiralities at the ends, respectively. Therefore, we have been able to demonstrate that, at least for nanotubes with high enough aspect ratio, the metastable states forming at intermediate stages of reversal can be completely tuned and controlled by the setting process previous to performance of the hysteresis loop. 


\subsection{Micromagnetic Simulations}

In order to further investigate the possibility to observe the above mentioned phenomenology in nanotubes with dimensions closer to usually synthesized nanotubes, which are in the range of tenths of nm, one has to revert to the micromagnetic approach instead of atomistic models. For this purpose, we have conducted  micromagnetic simulations of nanotubes having the same aspect ratio, namely $2.5$ as the (8,20) tube considered in the MC simulations, as described in Sec. \ref{Micromag_Sec} . 
At difference from the MC simulations, for which we compared loops started from different random seeds, here calculations were performed for two close values ($0.01$ and $0.03$) of the stopping $\left|m\times H\times m\right|$ torque parameter, measured in  A/m, which specifies that a stage should be considered complete when the torque across all spins drops below such value. 
It is worth to mentioning that the driver used for controlling minimization evolvers was \textbf{Oxs}\_\textbf{MinDriver} in the OOMMF package. 

\begin{figure}[htb]
\centering
\includegraphics[width=0.8\columnwidth]{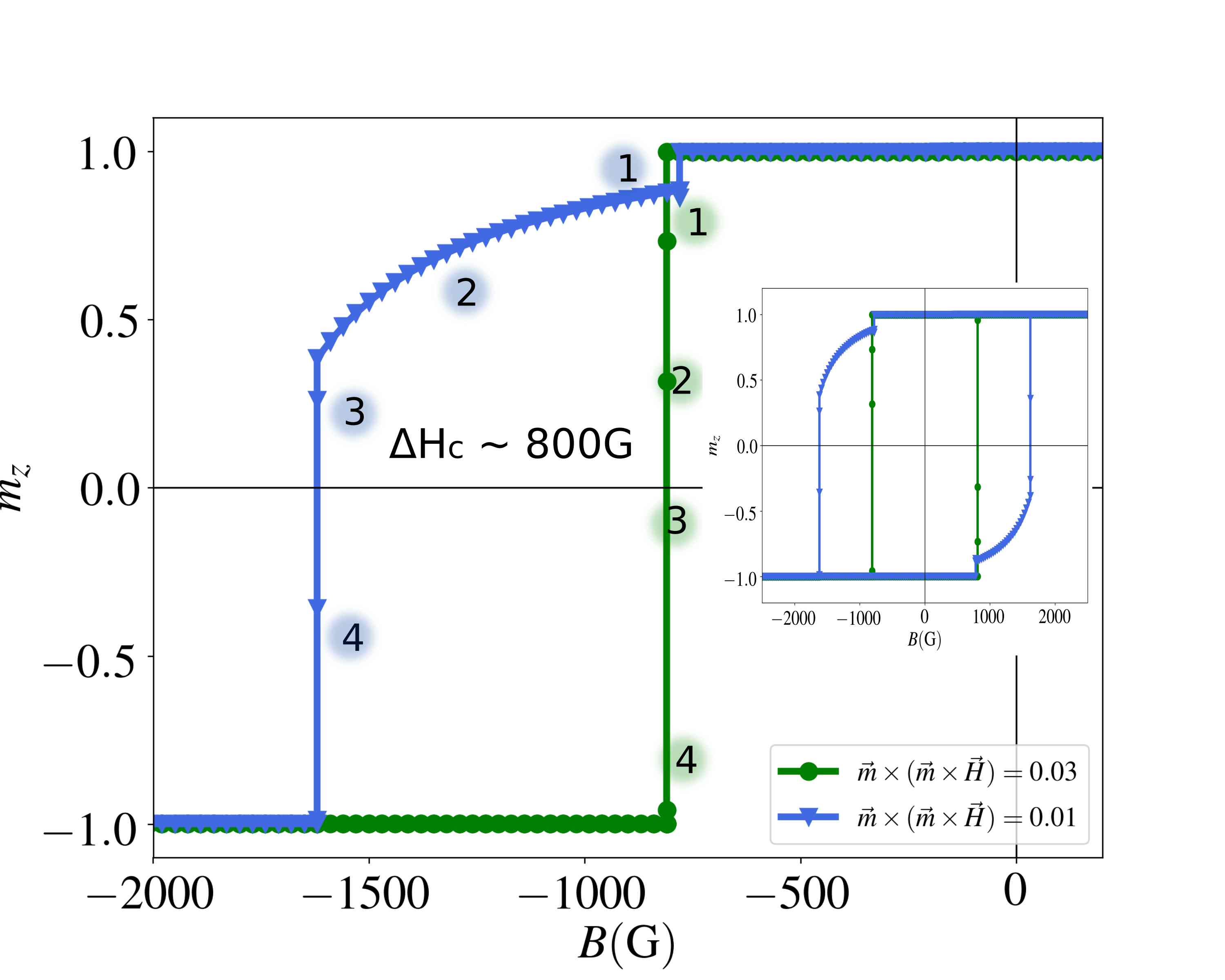}
\includegraphics[width=1.0\columnwidth]{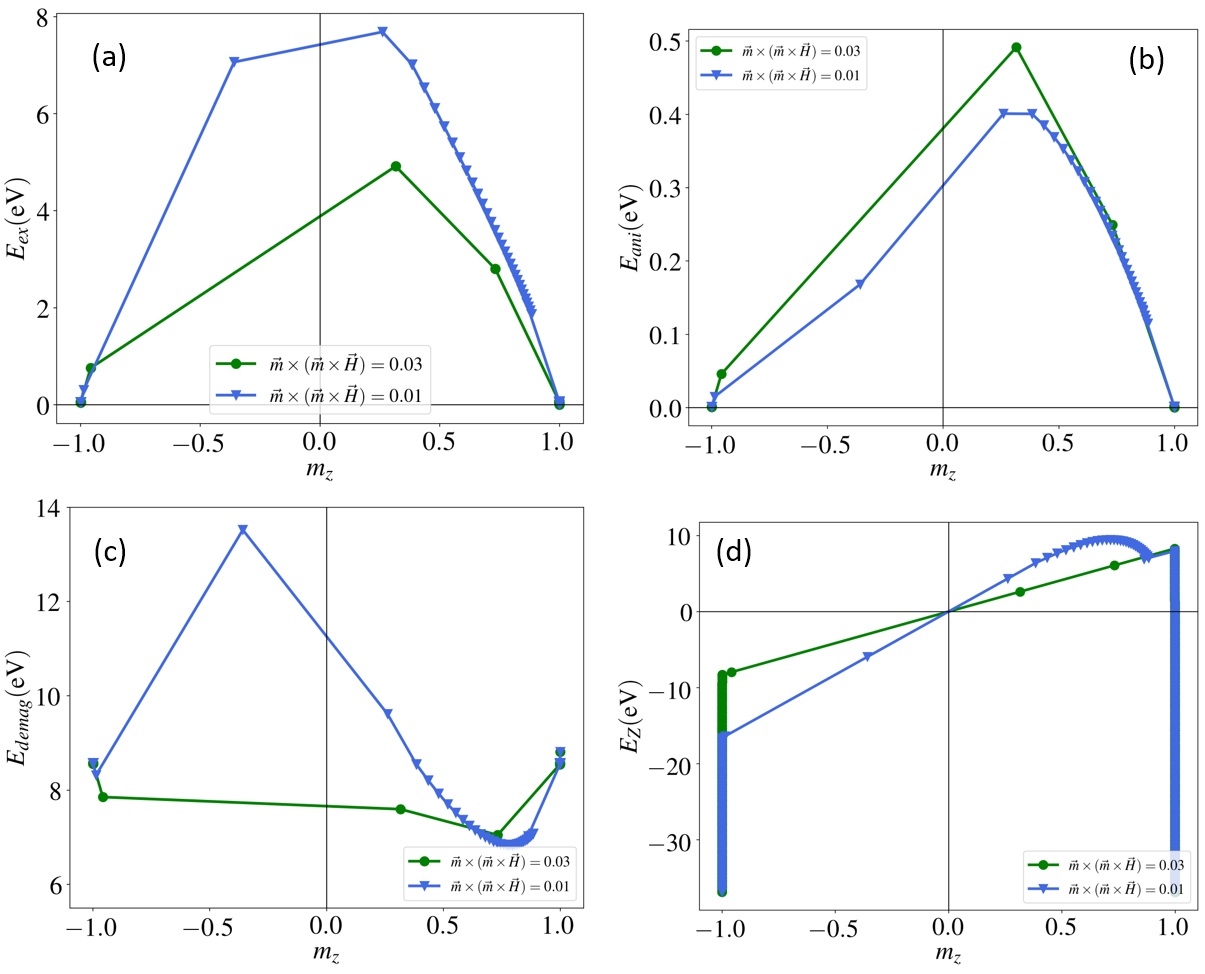}
\caption{Upper panel displays the decreasing field branches of the hysteresis loops simulated with OOMMF for a FeCo nanotube, starting from saturation, for two values of the magnetic torque in the stopping criteria. The inset shows the complete hysteresis loops. 
Lower panels display the dependence of the total (a), exchange (b), dipolar (c) and anisotropy (d) energies of the nanotube on its magnetization for the configurations attained along the decreasing field branch of the hysteresis loops shown in (a). 
}
\label{Hyst_oommf_Fig}
\end{figure}

\begin{figure}[htb]
\centering
\includegraphics[width=0.6\columnwidth]{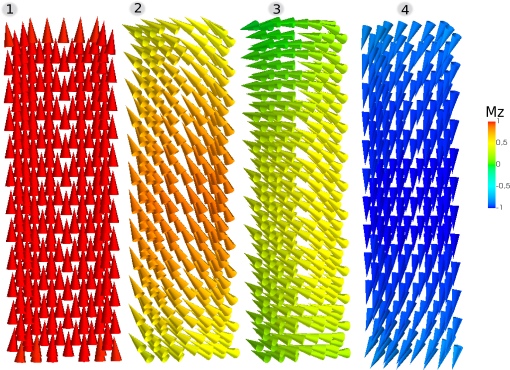}
\includegraphics[width=0.6\columnwidth]{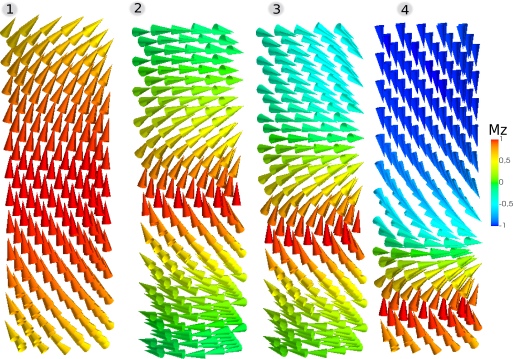}
\caption{Snapshots of the spin configurations in the reversal modes $Q1$ (upper panel) and Q2 (lower panel) of the hysteresis loops simulated . Numerical labels correspond to those shown in Fig. \ref{Hyst_oommf_Fig}.}
\label{Q1-Q2_confsoommf_Fig}
\end{figure}

The reason for choosing two different stopping parameters can be justified by the fact that macrostates along the curves of the hysteresis loops, although in thermal equilibrium, are not in mechanical equilibrium with the field, otherwise, no hysteresis would be observed. 
Typical values of this stopping parameter are in the range $0.1$ to $10$ and limits in the numerical precision of the energy calculations usually makes it not possible to obtain $\left|m\times H\times m\right|$ below about $0.01$ A/m.
As can be observed in Fig. \ref{Hyst_oommf_Fig}, two well-distinguished modes ($Q1$ and $Q2$) for magnetization reversal with two different coercivities were also obtained, analogously to the hysteresis loops obtained via MC for the (8,15) and (8,20) tubes.
In the same way, tracking of the spin configurations along the decreasing field branches of the loops allows to identify both switching modes in terms of a dual mechanism. The $Q1$ mode that follows an intermediate stage characterized by a vortices at the ends of the tube with the same chirality is obtained when the stopping torque is set to $0.03$, giving rise to a low coercive field. In contrast, when the stopping torque is set to $0.01$, the $Q2$ is observed, that is characterized by intermediate stages with vortices with opposite chirality that give rise to higher coercive field. Snapshots of these two reversal modes are shown in Fig. \ref{Q1-Q2_confsoommf_Fig}.
 
Likewise, the energy variation along the $m_z$-projection exhibits also a cross-point between two modes, indicating the inter-connectivity in phase space (see also Fig. 7 
in the Supplemental Material, where the energies dependence on the magnetic field are shown). The energies for the two cases are almost identical at high fields but they bifurcate at some negative field, where the mode with opposite chiralities makes an excursion to higher exchange energies before the tube magnetization is reversed. Similar jumps in energy are similar to those found, for example, during the reversal of vortices in nanodots  \cite{Scholz_jmmm2003}.


\section{Discussion and conclusions}
\label{Conclusions_Sec}

We have shown that the occurrence of both helical ($H$) and vortex ($V$) states during the magnetization reversal of nanotubes
is dictated by $\gamma$, the dipolar to exchange energy ratio. 
For a certain range of $\gamma$'s, the results of MC atomistic simulations have demonstrated that different reversal modes can occur along the hysteresis loop as a consequence of the high degree of metastability of the $H$ states that facilitates different paths through the energy landscape when varying the magnetic field.
In agreement with our results, recent experimental works on individual magnetic nanotubes \cite{Wyss_prb2017,Mehlin_prb2018} have also evidenced that short FeCoB nanotubes (0.6 $\mu$m long, $300$ mn in diameter), with similar aspect ratios as the ones studied here, can be found in mixed states with end vortices of opposing or matching circulations depending on the magnetic history or experiment repetition. Moreover, the comparison of cantilever magnetometry \cite{Mehlin_prb2018} with micromagnetic simulations showed that reversal initiated with matching vortices is correlated to lower energies and smoother energy variations than for opposing vortices.
It is worth mentioning also that in Ref. \onlinecite{Buchter_prl2013} some dispersion in the two branches of the hysteresis loops of Ni tubes was observed, which could be attributed to the different reversal possibilities shown here.

Our results allow to conclude that the reversals modes initiated by vortices with same chirality have coercive fields lower that the mode with opposite chiralities, since in this case the central region of the tube has to face the merging of the vortices due to the formation of a domain wall. The energy landscape suggests that both modes are interconnected in the energy phase space, giving rise to a high degree of metastability.
Our proposal to induce the reversal through the $Q_1$ or $Q_2$ modes, based on controlling the initial chirality at the tube ends, would allow to use a unique tube as a soft or hard material without changing its composition. In fact, the experimental study of Ref.\onlinecite{Mehlin_prb2018} have already suggested that control over relative chirality can be achieved introducing structural asymmetries at the nanotube ends, but we have shown that this might be achieved also without modifying the tube structure.

Although in this article we have limited our study to tubes of limited dimensions, we have given proof that our conclusions are not peculiar to the range of sizes studied. Preliminary results to be shown in a forthcoming publication, indicate that a similar phenomenology can be observed for longer or wider tubes, but in a range of $\gamma$'s that depend on the tube geometry. We also plan to study magnetization reversal under magnetic fields applied perpendicular to the tube axis and the include magnetocrystalline anisotropy into the simulations.

\acknowledgments
O. I. acknowledges financial support form the Spanish MINECO (MAT2015-68772), Catalan DURSI (2014SGR220) and European Union FEDER Funds (Una manera de hacer Europa), also CSUC for supercomputer facilities.
H.D.S and J.R acknowledge financial support from the Colciencias "Beca de Doctorados nacionales, convocatoria 727",  the project CODI-UdeA 2016-10085. J.R acknowledges UdeA for the exclusive dedication program.

\bibliography{hist_nanotubes_Oscar}

\begin{thebibliography}{56}%
\makeatletter
\providecommand \@ifxundefined [1]{%
 \@ifx{#1\undefined}
}%
\providecommand \@ifnum [1]{%
 \ifnum #1\expandafter \@firstoftwo
 \else \expandafter \@secondoftwo
 \fi
}%
\providecommand \@ifx [1]{%
 \ifx #1\expandafter \@firstoftwo
 \else \expandafter \@secondoftwo
 \fi
}%
\providecommand \natexlab [1]{#1}%
\providecommand \enquote  [1]{``#1''}%
\providecommand \bibnamefont  [1]{#1}%
\providecommand \bibfnamefont [1]{#1}%
\providecommand \citenamefont [1]{#1}%
\providecommand \href@noop [0]{\@secondoftwo}%
\providecommand \href [0]{\begingroup \@sanitize@url \@href}%
\providecommand \@href[1]{\@@startlink{#1}\@@href}%
\providecommand \@@href[1]{\endgroup#1\@@endlink}%
\providecommand \@sanitize@url [0]{\catcode `\\12\catcode `\$12\catcode
  `\&12\catcode `\#12\catcode `\^12\catcode `\_12\catcode `\%12\relax}%
\providecommand \@@startlink[1]{}%
\providecommand \@@endlink[0]{}%
\providecommand \url  [0]{\begingroup\@sanitize@url \@url }%
\providecommand \@url [1]{\endgroup\@href {#1}{\urlprefix }}%
\providecommand \urlprefix  [0]{URL }%
\providecommand \Eprint [0]{\href }%
\providecommand \doibase [0]{https://doi.org/}%
\providecommand \selectlanguage [0]{\@gobble}%
\providecommand \bibinfo  [0]{\@secondoftwo}%
\providecommand \bibfield  [0]{\@secondoftwo}%
\providecommand \translation [1]{[#1]}%
\providecommand \BibitemOpen [0]{}%
\providecommand \bibitemStop [0]{}%
\providecommand \bibitemNoStop [0]{.\EOS\space}%
\providecommand \EOS [0]{\spacefactor3000\relax}%
\providecommand \BibitemShut  [1]{\csname bibitem#1\endcsname}%
\let\auto@bib@innerbib\@empty
\bibitem [{\citenamefont {Staňo}\ and\ \citenamefont
  {Fruchart}(2018)}]{Stano_book2018}%
  \BibitemOpen
  \bibfield  {author} {\bibinfo {author} {\bibfnamefont {M.}~\bibnamefont
  {Staňo}}\ and\ \bibinfo {author} {\bibfnamefont {O.}~\bibnamefont
  {Fruchart}},\ }\bibfield  {title} {\bibinfo {title} {Chapter 3 - magnetic
  nanowires and nanotubes}\ }(\bibinfo  {publisher} {Elsevier},\ \bibinfo
  {year} {2018})\ pp.\ \bibinfo {pages} {155 -- 267}\BibitemShut {NoStop}%
\bibitem [{\citenamefont {Ye}\ and\ \citenamefont {Geng}(2012)}]{Ye_crssm2012}%
  \BibitemOpen
  \bibfield  {author} {\bibinfo {author} {\bibfnamefont {Y.}~\bibnamefont
  {Ye}}\ and\ \bibinfo {author} {\bibfnamefont {B.}~\bibnamefont {Geng}},\
  }\bibfield  {title} {\bibinfo {title} {{Magnetic Nanotubes: Synthesis,
  Properties, and Applications}},\ }\href
  {https://doi.org/10.1080/10408436.2011.613491} {\bibfield  {journal}
  {\bibinfo  {journal} {Critical Reviews in Solid State and Materials
  Sciences}\ }\textbf {\bibinfo {volume} {37}},\ \bibinfo {pages} {75}
  (\bibinfo {year} {2012})}\BibitemShut {NoStop}%
\bibitem [{\citenamefont {Cowburn}\ and\ \citenamefont
  {Welland}(2000)}]{Cowburn_science2000}%
  \BibitemOpen
  \bibfield  {author} {\bibinfo {author} {\bibfnamefont {R.~P.}\ \bibnamefont
  {Cowburn}}\ and\ \bibinfo {author} {\bibfnamefont {M.~E.}\ \bibnamefont
  {Welland}},\ }\bibfield  {title} {\bibinfo {title} {{Room temperature
  magnetic quantum cellular automata}},\ }\href
  {https://doi.org/10.1126/science.287.5457.1466} {\bibfield  {journal}
  {\bibinfo  {journal} {Science}\ }\textbf {\bibinfo {volume} {287}},\ \bibinfo
  {pages} {1466} (\bibinfo {year} {2000})}\BibitemShut {NoStop}%
\bibitem [{\citenamefont {Parkin}\ \emph {et~al.}(2008)\citenamefont {Parkin},
  \citenamefont {Hayashi},\ and\ \citenamefont {Thomas}}]{Parkin_science2008}%
  \BibitemOpen
  \bibfield  {author} {\bibinfo {author} {\bibfnamefont {S.~S.~P.}\
  \bibnamefont {Parkin}}, \bibinfo {author} {\bibfnamefont {M.}~\bibnamefont
  {Hayashi}},\ and\ \bibinfo {author} {\bibfnamefont {L.}~\bibnamefont
  {Thomas}},\ }\bibfield  {title} {\bibinfo {title} {Magnetic domain-wall
  racetrack memory},\ }\href {https://doi.org/10.1126/science.1145799}
  {\bibfield  {journal} {\bibinfo  {journal} {Science}\ }\textbf {\bibinfo
  {volume} {320}},\ \bibinfo {pages} {190} (\bibinfo {year}
  {2008})}\BibitemShut {NoStop}%
\bibitem [{\citenamefont {Alonso}\ \emph {et~al.}(2015)\citenamefont {Alonso},
  \citenamefont {Khurshid}, \citenamefont {Sankar}, \citenamefont {Nemati},
  \citenamefont {Phan}, \citenamefont {Garayo}, \citenamefont
  {Garc{\ifmmode\acute{\imath}\else\'{\i}\fi}a},\ and\ \citenamefont
  {Srikanth}}]{Alonso_jap2015}%
  \BibitemOpen
  \bibfield  {author} {\bibinfo {author} {\bibfnamefont {J.}~\bibnamefont
  {Alonso}}, \bibinfo {author} {\bibfnamefont {H.}~\bibnamefont {Khurshid}},
  \bibinfo {author} {\bibfnamefont {V.}~\bibnamefont {Sankar}}, \bibinfo
  {author} {\bibfnamefont {Z.}~\bibnamefont {Nemati}}, \bibinfo {author}
  {\bibfnamefont {M.~H.}\ \bibnamefont {Phan}}, \bibinfo {author}
  {\bibfnamefont {E.}~\bibnamefont {Garayo}}, \bibinfo {author} {\bibfnamefont
  {J.~A.}\ \bibnamefont {Garc{\ifmmode\acute{\imath}\else\'{\i}\fi}a}},\ and\
  \bibinfo {author} {\bibfnamefont {H.}~\bibnamefont {Srikanth}},\ }\bibfield
  {title} {\bibinfo {title} {{FeCo nanowires with enhanced heating powers and
  controllable dimensions for magnetic hyperthermia}},\ }\href
  {https://doi.org/10.1063/1.4908300} {\bibfield  {journal} {\bibinfo
  {journal} {J. Appl. Phys.}\ }\textbf {\bibinfo {volume} {117}},\ \bibinfo
  {pages} {17D113} (\bibinfo {year} {2015})}\BibitemShut {NoStop}%
\bibitem [{\citenamefont {Fernandez-Roldan}\ \emph
  {et~al.}(2018{\natexlab{a}})\citenamefont {Fernandez-Roldan}, \citenamefont
  {Serantes}, \citenamefont {del Real}, \citenamefont {Vazquez},\ and\
  \citenamefont {Chubykalo-Fesenko}}]{Fernandez-Roldan_apl2018}%
  \BibitemOpen
  \bibfield  {author} {\bibinfo {author} {\bibfnamefont {J.~A.}\ \bibnamefont
  {Fernandez-Roldan}}, \bibinfo {author} {\bibfnamefont {D.}~\bibnamefont
  {Serantes}}, \bibinfo {author} {\bibfnamefont {R.~P.}\ \bibnamefont {del
  Real}}, \bibinfo {author} {\bibfnamefont {M.}~\bibnamefont {Vazquez}},\ and\
  \bibinfo {author} {\bibfnamefont {O.}~\bibnamefont {Chubykalo-Fesenko}},\
  }\bibfield  {title} {\bibinfo {title} {{Micromagnetic evaluation of the
  dissipated heat in cylindrical magnetic nanowires}},\ }\href
  {https://doi.org/10.1063/1.5025922} {\bibfield  {journal} {\bibinfo
  {journal} {Appl. Phys. Lett.}\ }\textbf {\bibinfo {volume} {112}},\ \bibinfo
  {pages} {212402} (\bibinfo {year} {2018}{\natexlab{a}})}\BibitemShut
  {NoStop}%
\bibitem [{\citenamefont {Chen}\ \emph {et~al.}(2012)\citenamefont {Chen},
  \citenamefont {Klingeler}, \citenamefont {Kath}, \citenamefont {Gendy},
  \citenamefont {Cendrowski}, \citenamefont {Kalenczuk},\ and\ \citenamefont
  {Borowiak-palen}}]{Chen_acs2012}%
  \BibitemOpen
  \bibfield  {author} {\bibinfo {author} {\bibfnamefont {X.}~\bibnamefont
  {Chen}}, \bibinfo {author} {\bibfnamefont {R.}~\bibnamefont {Klingeler}},
  \bibinfo {author} {\bibfnamefont {M.}~\bibnamefont {Kath}}, \bibinfo {author}
  {\bibfnamefont {A.~A.~E.}\ \bibnamefont {Gendy}}, \bibinfo {author}
  {\bibfnamefont {K.}~\bibnamefont {Cendrowski}}, \bibinfo {author}
  {\bibfnamefont {R.~J.}\ \bibnamefont {Kalenczuk}},\ and\ \bibinfo {author}
  {\bibfnamefont {E.}~\bibnamefont {Borowiak-palen}},\ }\bibfield  {title}
  {\bibinfo {title} {{Magnetic Silica Nanotubes: Synthesis, Drug Release, and
  Feasibility for Magnetic Hyperthermia}},\ }\href
  {https://doi.org/10.1021/am300469r} {\bibfield  {journal} {\bibinfo
  {journal} {ACS Applied Materials \& Interfaces}\ }\textbf {\bibinfo {volume}
  {4}},\ \bibinfo {pages} {2303} (\bibinfo {year} {2012})}\BibitemShut
  {NoStop}%
\bibitem [{\citenamefont {Salinas}\ and\ \citenamefont
  {Restrepo}(2012)}]{Salinas_jsnm2012}%
  \BibitemOpen
  \bibfield  {author} {\bibinfo {author} {\bibfnamefont {H.~D.}\ \bibnamefont
  {Salinas}}\ and\ \bibinfo {author} {\bibfnamefont {J.}~\bibnamefont
  {Restrepo}},\ }\bibfield  {title} {\bibinfo {title} {Influence of the
  competition between dipolar and exchange interactions on the magnetic
  structure of single-wall nanocylinders. monte carlo simulation},\ }\href
  {https://doi.org/10.1007/s10948-012-1651-9} {\bibfield  {journal} {\bibinfo
  {journal} {J. Supercond. Nov. Magn.}\ }\textbf {\bibinfo {volume} {25}},\
  \bibinfo {pages} {2217} (\bibinfo {year} {2012})}\BibitemShut {NoStop}%
\bibitem [{\citenamefont {Chen}\ \emph {et~al.}(2007)\citenamefont {Chen},
  \citenamefont {Usov}, \citenamefont {Blanco},\ and\ \citenamefont
  {Gonzalez}}]{Chen_jmmm2007}%
  \BibitemOpen
  \bibfield  {author} {\bibinfo {author} {\bibfnamefont {A.}~\bibnamefont
  {Chen}}, \bibinfo {author} {\bibfnamefont {N.}~\bibnamefont {Usov}}, \bibinfo
  {author} {\bibfnamefont {J.}~\bibnamefont {Blanco}},\ and\ \bibinfo {author}
  {\bibfnamefont {J.}~\bibnamefont {Gonzalez}},\ }\bibfield  {title} {\bibinfo
  {title} {Equilibrium magnetization states in magnetic nanotubes and their
  evolution in external magnetic field},\ }\href
  {https://doi.org/10.1016/j.jmmm.2007.02.132} {\bibfield  {journal} {\bibinfo
  {journal} {J. Magn. Magn. Mater.}\ }\textbf {\bibinfo {volume} {316}},\
  \bibinfo {pages} {e317 } (\bibinfo {year} {2007})}\BibitemShut {NoStop}%
\bibitem [{\citenamefont {Chen}\ \emph {et~al.}(2010)\citenamefont {Chen},
  \citenamefont {Guslienko},\ and\ \citenamefont {Gonzalez}}]{Chen_jap2010}%
  \BibitemOpen
  \bibfield  {author} {\bibinfo {author} {\bibfnamefont {A.~P.}\ \bibnamefont
  {Chen}}, \bibinfo {author} {\bibfnamefont {K.~Y.}\ \bibnamefont
  {Guslienko}},\ and\ \bibinfo {author} {\bibfnamefont {J.}~\bibnamefont
  {Gonzalez}},\ }\bibfield  {title} {\bibinfo {title} {Magnetization
  configurations and reversal of thin magnetic nanotubes with uniaxial
  anisotropy},\ }\href {https://doi.org/10.1063/1.3488630} {\bibfield
  {journal} {\bibinfo  {journal} {J. Appl. Phys.}\ }\textbf {\bibinfo {volume}
  {108}},\ \bibinfo {pages} {083920} (\bibinfo {year} {2010})}\BibitemShut
  {NoStop}%
\bibitem [{\citenamefont {Chen}\ \emph {et~al.}(2011)\citenamefont {Chen},
  \citenamefont {Gonzalez},\ and\ \citenamefont {Guslienko}}]{Chen_jap2011}%
  \BibitemOpen
  \bibfield  {author} {\bibinfo {author} {\bibfnamefont {A.-P.}\ \bibnamefont
  {Chen}}, \bibinfo {author} {\bibfnamefont {J.~M.}\ \bibnamefont {Gonzalez}},\
  and\ \bibinfo {author} {\bibfnamefont {K.~Y.}\ \bibnamefont {Guslienko}},\
  }\bibfield  {title} {\bibinfo {title} {Magnetization configurations and
  reversal of magnetic nanotubes with opposite chiralities of the end
  domains},\ }\href {https://doi.org/10.1063/1.3562190} {\bibfield  {journal}
  {\bibinfo  {journal} {J. Appl. Phys.}\ }\textbf {\bibinfo {volume} {109}},\
  \bibinfo {pages} {073923} (\bibinfo {year} {2011})}\BibitemShut {NoStop}%
\bibitem [{\citenamefont {Escrig}\ \emph {et~al.}(2007)\citenamefont {Escrig},
  \citenamefont {Landeros}, \citenamefont {Altbir}, \citenamefont {Vogel},\
  and\ \citenamefont {Vargas}}]{Escrig_jmmm2007}%
  \BibitemOpen
  \bibfield  {author} {\bibinfo {author} {\bibfnamefont {J.}~\bibnamefont
  {Escrig}}, \bibinfo {author} {\bibfnamefont {P.}~\bibnamefont {Landeros}},
  \bibinfo {author} {\bibfnamefont {D.}~\bibnamefont {Altbir}}, \bibinfo
  {author} {\bibfnamefont {E.}~\bibnamefont {Vogel}},\ and\ \bibinfo {author}
  {\bibfnamefont {P.}~\bibnamefont {Vargas}},\ }\bibfield  {title} {\bibinfo
  {title} {Phase diagrams of magnetic nanotubes},\ }\href
  {https://doi.org/10.1016/j.jmmm.2006.05.019} {\bibfield  {journal} {\bibinfo
  {journal} {J. Magn. Magn. Mater.}\ }\textbf {\bibinfo {volume} {308}},\
  \bibinfo {pages} {233 } (\bibinfo {year} {2007})}\BibitemShut {NoStop}%
\bibitem [{\citenamefont {Biziere}\ \emph {et~al.}(2013)\citenamefont
  {Biziere}, \citenamefont {Gatel}, \citenamefont {Lassalle-Balier},
  \citenamefont {Clochard}, \citenamefont {Wegrowe},\ and\ \citenamefont
  {Snoeck}}]{Biziere_Nanolett2013}%
  \BibitemOpen
  \bibfield  {author} {\bibinfo {author} {\bibfnamefont {N.}~\bibnamefont
  {Biziere}}, \bibinfo {author} {\bibfnamefont {C.}~\bibnamefont {Gatel}},
  \bibinfo {author} {\bibfnamefont {R.}~\bibnamefont {Lassalle-Balier}},
  \bibinfo {author} {\bibfnamefont {M.~C.}\ \bibnamefont {Clochard}}, \bibinfo
  {author} {\bibfnamefont {J.~E.}\ \bibnamefont {Wegrowe}},\ and\ \bibinfo
  {author} {\bibfnamefont {E.}~\bibnamefont {Snoeck}},\ }\bibfield  {title}
  {\bibinfo {title} {Imaging the fine structure of a magnetic domain wall in a
  ni nanocylinder},\ }\href {https://doi.org/10.1021/nl400317j} {\bibfield
  {journal} {\bibinfo  {journal} {Nano Letters}\ }\textbf {\bibinfo {volume}
  {13}},\ \bibinfo {pages} {2053} (\bibinfo {year} {2013})}\BibitemShut
  {NoStop}%
\bibitem [{\citenamefont {Weber}\ \emph {et~al.}(2012)\citenamefont {Weber},
  \citenamefont {Rüffer}, \citenamefont {Buchter}, \citenamefont {Xue},
  \citenamefont {Russo-Averchi}, \citenamefont {Huber}, \citenamefont
  {Berberich}, \citenamefont {Arbiol}, \citenamefont {Fontcuberta~i Morral},
  \citenamefont {Grundler},\ and\ \citenamefont {Poggio}}]{Weber_Nanolett2012}%
  \BibitemOpen
  \bibfield  {author} {\bibinfo {author} {\bibfnamefont {D.~P.}\ \bibnamefont
  {Weber}}, \bibinfo {author} {\bibfnamefont {D.}~\bibnamefont {Rüffer}},
  \bibinfo {author} {\bibfnamefont {A.}~\bibnamefont {Buchter}}, \bibinfo
  {author} {\bibfnamefont {F.}~\bibnamefont {Xue}}, \bibinfo {author}
  {\bibfnamefont {E.}~\bibnamefont {Russo-Averchi}}, \bibinfo {author}
  {\bibfnamefont {R.}~\bibnamefont {Huber}}, \bibinfo {author} {\bibfnamefont
  {P.}~\bibnamefont {Berberich}}, \bibinfo {author} {\bibfnamefont
  {J.}~\bibnamefont {Arbiol}}, \bibinfo {author} {\bibfnamefont
  {A.}~\bibnamefont {Fontcuberta~i Morral}}, \bibinfo {author} {\bibfnamefont
  {D.}~\bibnamefont {Grundler}},\ and\ \bibinfo {author} {\bibfnamefont
  {M.}~\bibnamefont {Poggio}},\ }\bibfield  {title} {\bibinfo {title}
  {Cantilever magnetometry of individual ni nanotubes},\ }\href
  {https://doi.org/10.1021/nl302950u} {\bibfield  {journal} {\bibinfo
  {journal} {Nano Letters}\ }\textbf {\bibinfo {volume} {12}},\ \bibinfo
  {pages} {6139} (\bibinfo {year} {2012})}\BibitemShut {NoStop}%
\bibitem [{\citenamefont {Buchter}\ \emph {et~al.}(2013)\citenamefont
  {Buchter}, \citenamefont {Nagel}, \citenamefont {R\"uffer}, \citenamefont
  {Xue}, \citenamefont {Weber}, \citenamefont {Kieler}, \citenamefont
  {Weimann}, \citenamefont {Kohlmann}, \citenamefont {Zorin}, \citenamefont
  {Russo-Averchi}, \citenamefont {Huber}, \citenamefont {Berberich},
  \citenamefont {Fontcuberta~i Morral}, \citenamefont {Kemmler}, \citenamefont
  {Kleiner}, \citenamefont {Koelle}, \citenamefont {Grundler},\ and\
  \citenamefont {Poggio}}]{Buchter_prl2013}%
  \BibitemOpen
  \bibfield  {author} {\bibinfo {author} {\bibfnamefont {A.}~\bibnamefont
  {Buchter}}, \bibinfo {author} {\bibfnamefont {J.}~\bibnamefont {Nagel}},
  \bibinfo {author} {\bibfnamefont {D.}~\bibnamefont {R\"uffer}}, \bibinfo
  {author} {\bibfnamefont {F.}~\bibnamefont {Xue}}, \bibinfo {author}
  {\bibfnamefont {D.~P.}\ \bibnamefont {Weber}}, \bibinfo {author}
  {\bibfnamefont {O.~F.}\ \bibnamefont {Kieler}}, \bibinfo {author}
  {\bibfnamefont {T.}~\bibnamefont {Weimann}}, \bibinfo {author} {\bibfnamefont
  {J.}~\bibnamefont {Kohlmann}}, \bibinfo {author} {\bibfnamefont {A.~B.}\
  \bibnamefont {Zorin}}, \bibinfo {author} {\bibfnamefont {E.}~\bibnamefont
  {Russo-Averchi}}, \bibinfo {author} {\bibfnamefont {R.}~\bibnamefont
  {Huber}}, \bibinfo {author} {\bibfnamefont {P.}~\bibnamefont {Berberich}},
  \bibinfo {author} {\bibfnamefont {A.}~\bibnamefont {Fontcuberta~i Morral}},
  \bibinfo {author} {\bibfnamefont {M.}~\bibnamefont {Kemmler}}, \bibinfo
  {author} {\bibfnamefont {R.}~\bibnamefont {Kleiner}}, \bibinfo {author}
  {\bibfnamefont {D.}~\bibnamefont {Koelle}}, \bibinfo {author} {\bibfnamefont
  {D.}~\bibnamefont {Grundler}},\ and\ \bibinfo {author} {\bibfnamefont
  {M.}~\bibnamefont {Poggio}},\ }\bibfield  {title} {\bibinfo {title} {Reversal
  mechanism of an individual ni nanotube simultaneously studied by torque and
  squid magnetometry},\ }\href {https://doi.org/10.1103/PhysRevLett.111.067202}
  {\bibfield  {journal} {\bibinfo  {journal} {Phys. Rev. Lett.}\ }\textbf
  {\bibinfo {volume} {111}},\ \bibinfo {pages} {067202} (\bibinfo {year}
  {2013})}\BibitemShut {NoStop}%
\bibitem [{\citenamefont {Yamasaki}\ \emph {et~al.}(2003)\citenamefont
  {Yamasaki}, \citenamefont {Wulfhekel}, \citenamefont {Hertel}, \citenamefont
  {Suga},\ and\ \citenamefont {Kirschner}}]{Yamasaki_prl2003}%
  \BibitemOpen
  \bibfield  {author} {\bibinfo {author} {\bibfnamefont {A.}~\bibnamefont
  {Yamasaki}}, \bibinfo {author} {\bibfnamefont {W.}~\bibnamefont {Wulfhekel}},
  \bibinfo {author} {\bibfnamefont {R.}~\bibnamefont {Hertel}}, \bibinfo
  {author} {\bibfnamefont {S.}~\bibnamefont {Suga}},\ and\ \bibinfo {author}
  {\bibfnamefont {J.}~\bibnamefont {Kirschner}},\ }\bibfield  {title} {\bibinfo
  {title} {Direct observation of the single-domain limit of fe nanomagnets by
  spin-polarized scanning tunneling spectroscopy},\ }\href
  {https://doi.org/10.1103/PhysRevLett.91.127201} {\bibfield  {journal}
  {\bibinfo  {journal} {Phys. Rev. Lett.}\ }\textbf {\bibinfo {volume} {91}},\
  \bibinfo {pages} {127201} (\bibinfo {year} {2003})}\BibitemShut {NoStop}%
\bibitem [{\citenamefont {Salinas}\ \emph {et~al.}(2018)\citenamefont
  {Salinas}, \citenamefont {Restrepo},\ and\ \citenamefont
  {Iglesias}}]{Salinas_scirep2018}%
  \BibitemOpen
  \bibfield  {author} {\bibinfo {author} {\bibfnamefont {H.~D.}\ \bibnamefont
  {Salinas}}, \bibinfo {author} {\bibfnamefont {J.}~\bibnamefont {Restrepo}},\
  and\ \bibinfo {author} {\bibfnamefont {{\`{O}}.}~\bibnamefont {Iglesias}},\
  }\bibfield  {title} {\bibinfo {title} {{Change in the magnetic configurations
  of tubular nanostructures by tuning dipolar interactions}},\ }\href
  {https://doi.org/10.1038/s41598-018-28598-1} {\bibfield  {journal} {\bibinfo
  {journal} {Scientific Reports}\ }\textbf {\bibinfo {volume} {8}},\ \bibinfo
  {pages} {10275} (\bibinfo {year} {2018})}\BibitemShut {NoStop}%
\bibitem [{\citenamefont {Proenca}\ \emph {et~al.}(2013)\citenamefont
  {Proenca}, \citenamefont {Sousa}, \citenamefont {Escrig}, \citenamefont
  {Ventura}, \citenamefont {Vazquez},\ and\ \citenamefont
  {Araujo}}]{Proenca_jap2013}%
  \BibitemOpen
  \bibfield  {author} {\bibinfo {author} {\bibfnamefont {M.~P.}\ \bibnamefont
  {Proenca}}, \bibinfo {author} {\bibfnamefont {C.~T.}\ \bibnamefont {Sousa}},
  \bibinfo {author} {\bibfnamefont {J.}~\bibnamefont {Escrig}}, \bibinfo
  {author} {\bibfnamefont {J.}~\bibnamefont {Ventura}}, \bibinfo {author}
  {\bibfnamefont {M.}~\bibnamefont {Vazquez}},\ and\ \bibinfo {author}
  {\bibfnamefont {J.~P.}\ \bibnamefont {Araujo}},\ }\bibfield  {title}
  {\bibinfo {title} {{Magnetic interactions and reversal mechanisms in Co
  nanowire and nanotube arrays}},\ }\href {https://doi.org/10.1063/1.4794335}
  {\bibfield  {journal} {\bibinfo  {journal} {J. Appl. Phys.}\ }\textbf
  {\bibinfo {volume} {113}},\ \bibinfo {pages} {093907} (\bibinfo {year}
  {2013})}\BibitemShut {NoStop}%
\bibitem [{\citenamefont {Nielsch}\ \emph {et~al.}(2005)\citenamefont
  {Nielsch}, \citenamefont {Castaño}, \citenamefont {Ross},\ and\
  \citenamefont {Krishnan}}]{Nielsch_jap2005}%
  \BibitemOpen
  \bibfield  {author} {\bibinfo {author} {\bibfnamefont {K.}~\bibnamefont
  {Nielsch}}, \bibinfo {author} {\bibfnamefont {F.~J.}\ \bibnamefont
  {Castaño}}, \bibinfo {author} {\bibfnamefont {C.~A.}\ \bibnamefont {Ross}},\
  and\ \bibinfo {author} {\bibfnamefont {R.}~\bibnamefont {Krishnan}},\
  }\bibfield  {title} {\bibinfo {title} {Magnetic properties of
  template-synthesized cobalt polymer composite nanotubes},\ }\bibfield
  {journal} {\bibinfo  {journal} {J. Appl. Phys.}\ }\textbf {\bibinfo {volume}
  {98}},\ \href {https://doi.org/10.1063/1.2005384} {10.1063/1.2005384}
  (\bibinfo {year} {2005})\BibitemShut {NoStop}%
\bibitem [{\citenamefont {Han}\ \emph {et~al.}(2009)\citenamefont {Han},
  \citenamefont {Shamaila}, \citenamefont {Sharif}, \citenamefont {Chen},
  \citenamefont {Liu},\ and\ \citenamefont {Liu}}]{Han_advmat2009}%
  \BibitemOpen
  \bibfield  {author} {\bibinfo {author} {\bibfnamefont {X.~F.}\ \bibnamefont
  {Han}}, \bibinfo {author} {\bibfnamefont {S.}~\bibnamefont {Shamaila}},
  \bibinfo {author} {\bibfnamefont {R.}~\bibnamefont {Sharif}}, \bibinfo
  {author} {\bibfnamefont {J.~Y.}\ \bibnamefont {Chen}}, \bibinfo {author}
  {\bibfnamefont {H.~R.}\ \bibnamefont {Liu}},\ and\ \bibinfo {author}
  {\bibfnamefont {D.~P.}\ \bibnamefont {Liu}},\ }\bibfield  {title} {\bibinfo
  {title} {{Structural and magnetic properties of various ferromagnetic
  nanotubes}},\ }\href {https://doi.org/10.1002/adma.200901065} {\bibfield
  {journal} {\bibinfo  {journal} {Advanced Materials}\ }\textbf {\bibinfo
  {volume} {21}},\ \bibinfo {pages} {4619} (\bibinfo {year}
  {2009})}\BibitemShut {NoStop}%
\bibitem [{\citenamefont {Sui}\ \emph {et~al.}(2004)\citenamefont {Sui},
  \citenamefont {Skomski}, \citenamefont {Sorge},\ and\ \citenamefont
  {Sellmyer}}]{Sui_apl2004}%
  \BibitemOpen
  \bibfield  {author} {\bibinfo {author} {\bibfnamefont {Y.~C.}\ \bibnamefont
  {Sui}}, \bibinfo {author} {\bibfnamefont {R.}~\bibnamefont {Skomski}},
  \bibinfo {author} {\bibfnamefont {K.~D.}\ \bibnamefont {Sorge}},\ and\
  \bibinfo {author} {\bibfnamefont {D.~J.}\ \bibnamefont {Sellmyer}},\
  }\bibfield  {title} {\bibinfo {title} {Nanotube magnetism},\ }\href
  {https://doi.org/10.1063/1.1655692} {\bibfield  {journal} {\bibinfo
  {journal} {Appl. Phys. Lett.}\ }\textbf {\bibinfo {volume} {84}},\ \bibinfo
  {pages} {1525} (\bibinfo {year} {2004})}\BibitemShut {NoStop}%
\bibitem [{\citenamefont {Xu}\ \emph {et~al.}(2008)\citenamefont {Xu},
  \citenamefont {Xue}, \citenamefont {Fu}, \citenamefont {Gao},\ and\
  \citenamefont {Gao}}]{Xu_jpd2008}%
  \BibitemOpen
  \bibfield  {author} {\bibinfo {author} {\bibfnamefont {Y.}~\bibnamefont
  {Xu}}, \bibinfo {author} {\bibfnamefont {D.~S.}\ \bibnamefont {Xue}},
  \bibinfo {author} {\bibfnamefont {J.~L.}\ \bibnamefont {Fu}}, \bibinfo
  {author} {\bibfnamefont {D.~Q.}\ \bibnamefont {Gao}},\ and\ \bibinfo {author}
  {\bibfnamefont {B.}~\bibnamefont {Gao}},\ }\bibfield  {title} {\bibinfo
  {title} {Synthesis, characterization and magnetic properties of {Fe}
  nanotubes},\ }\href {https://doi.org/10.1088/0022-3727/41/21/215010}
  {\bibfield  {journal} {\bibinfo  {journal} {J. Phys. D: Appl. Phys.}\
  }\textbf {\bibinfo {volume} {41}},\ \bibinfo {pages} {215010} (\bibinfo
  {year} {2008})}\BibitemShut {NoStop}%
\bibitem [{\citenamefont {Daub}\ \emph {et~al.}(2007)\citenamefont {Daub},
  \citenamefont {Knez}, \citenamefont {Goesele},\ and\ \citenamefont
  {Nielsch}}]{Daub_jap2007}%
  \BibitemOpen
  \bibfield  {author} {\bibinfo {author} {\bibfnamefont {M.}~\bibnamefont
  {Daub}}, \bibinfo {author} {\bibfnamefont {M.}~\bibnamefont {Knez}}, \bibinfo
  {author} {\bibfnamefont {U.}~\bibnamefont {Goesele}},\ and\ \bibinfo {author}
  {\bibfnamefont {K.}~\bibnamefont {Nielsch}},\ }\bibfield  {title} {\bibinfo
  {title} {{Ferromagnetic nanotubes by atomic layer deposition in anodic
  alumina membranes}},\ }\href {https://doi.org/10.1063/1.2712057} {\bibfield
  {journal} {\bibinfo  {journal} {J. Appl. Phys.}\ }\textbf {\bibinfo {volume}
  {101}},\ \bibinfo {pages} {1} (\bibinfo {year} {2007})}\BibitemShut {NoStop}%
\bibitem [{\citenamefont {Garcia}\ \emph {et~al.}(2018)\citenamefont {Garcia},
  \citenamefont {Rosa}, \citenamefont {Garcia}, \citenamefont {Prida},
  \citenamefont {Hernando}, \citenamefont {L{\'{o}}pez}, \citenamefont
  {Vargas},\ and\ \citenamefont {Ross}}]{Garcia_jpc2018}%
  \BibitemOpen
  \bibfield  {author} {\bibinfo {author} {\bibfnamefont {C.}~\bibnamefont
  {Garcia}}, \bibinfo {author} {\bibfnamefont {W.~O.}\ \bibnamefont {Rosa}},
  \bibinfo {author} {\bibfnamefont {J.}~\bibnamefont {Garcia}}, \bibinfo
  {author} {\bibfnamefont {V.~M.}\ \bibnamefont {Prida}}, \bibinfo {author}
  {\bibfnamefont {B.}~\bibnamefont {Hernando}}, \bibinfo {author}
  {\bibfnamefont {J.~A.}\ \bibnamefont {L{\'{o}}pez}}, \bibinfo {author}
  {\bibfnamefont {P.}~\bibnamefont {Vargas}},\ and\ \bibinfo {author}
  {\bibfnamefont {C.~A.}\ \bibnamefont {Ross}},\ }\bibfield  {title} {\bibinfo
  {title} {{Magnetization Reversal in Radially Distributed Nanowire Arrays}},\
  }\href {https://doi.org/10.1021/acs.jpcc.7b10799} {\bibfield  {journal}
  {\bibinfo  {journal} {Journal of Physical Chemistry C}\ }\textbf {\bibinfo
  {volume} {122}},\ \bibinfo {pages} {5124} (\bibinfo {year}
  {2018})}\BibitemShut {NoStop}%
\bibitem [{\citenamefont {Streubel}\ \emph {et~al.}(2015)\citenamefont
  {Streubel}, \citenamefont {Kronast}, \citenamefont {Fischer}, \citenamefont
  {Parkinson}, \citenamefont {Schmidt},\ and\ \citenamefont
  {Makarov}}]{Streubel_natcomm2015}%
  \BibitemOpen
  \bibfield  {author} {\bibinfo {author} {\bibfnamefont {R.}~\bibnamefont
  {Streubel}}, \bibinfo {author} {\bibfnamefont {F.}~\bibnamefont {Kronast}},
  \bibinfo {author} {\bibfnamefont {P.}~\bibnamefont {Fischer}}, \bibinfo
  {author} {\bibfnamefont {D.}~\bibnamefont {Parkinson}}, \bibinfo {author}
  {\bibfnamefont {O.~G.}\ \bibnamefont {Schmidt}},\ and\ \bibinfo {author}
  {\bibfnamefont {D.}~\bibnamefont {Makarov}},\ }\bibfield  {title} {\bibinfo
  {title} {Retrieving spin textures on curved magnetic thin films with
  full-field soft x-ray microscopies},\ }\href
  {https://doi.org/10.1038/ncomms8612} {\bibfield  {journal} {\bibinfo
  {journal} {Nature Communications}\ }\textbf {\bibinfo {volume} {6}},\
  \bibinfo {pages} {7612} (\bibinfo {year} {2015})}\BibitemShut {NoStop}%
\bibitem [{\citenamefont {Donnelly}\ \emph {et~al.}(2017)\citenamefont
  {Donnelly}, \citenamefont {Guizar-Sicairos}, \citenamefont {Scagnoli},
  \citenamefont {Gliga}, \citenamefont {Holler}, \citenamefont {Raabe},\ and\
  \citenamefont {Heyderman}}]{Donnelly_nature2017}%
  \BibitemOpen
  \bibfield  {author} {\bibinfo {author} {\bibfnamefont {C.}~\bibnamefont
  {Donnelly}}, \bibinfo {author} {\bibfnamefont {M.}~\bibnamefont
  {Guizar-Sicairos}}, \bibinfo {author} {\bibfnamefont {V.}~\bibnamefont
  {Scagnoli}}, \bibinfo {author} {\bibfnamefont {S.}~\bibnamefont {Gliga}},
  \bibinfo {author} {\bibfnamefont {M.}~\bibnamefont {Holler}}, \bibinfo
  {author} {\bibfnamefont {J.}~\bibnamefont {Raabe}},\ and\ \bibinfo {author}
  {\bibfnamefont {L.~J.}\ \bibnamefont {Heyderman}},\ }\bibfield  {title}
  {\bibinfo {title} {{Three-dimensional magnetization structures revealed with
  X-ray vector nanotomography}},\ }\href {https://doi.org/10.1038/nature23006}
  {\bibfield  {journal} {\bibinfo  {journal} {Nature}\ }\textbf {\bibinfo
  {volume} {547}},\ \bibinfo {pages} {328} (\bibinfo {year}
  {2017})}\BibitemShut {NoStop}%
\bibitem [{\citenamefont {Zimmermann}\ \emph {et~al.}(2018)\citenamefont
  {Zimmermann}, \citenamefont {Meier}, \citenamefont {Dirnberger},
  \citenamefont {Kákay}, \citenamefont {Decker}, \citenamefont {Wintz},
  \citenamefont {Finizio}, \citenamefont {Josten}, \citenamefont {Raabe},
  \citenamefont {Kronseder}, \citenamefont {Bougeard}, \citenamefont
  {Lindner},\ and\ \citenamefont {Back}}]{Zimmermann_nanol2018}%
  \BibitemOpen
  \bibfield  {author} {\bibinfo {author} {\bibfnamefont {M.}~\bibnamefont
  {Zimmermann}}, \bibinfo {author} {\bibfnamefont {T.~N.~G.}\ \bibnamefont
  {Meier}}, \bibinfo {author} {\bibfnamefont {F.}~\bibnamefont {Dirnberger}},
  \bibinfo {author} {\bibfnamefont {A.}~\bibnamefont {Kákay}}, \bibinfo
  {author} {\bibfnamefont {M.}~\bibnamefont {Decker}}, \bibinfo {author}
  {\bibfnamefont {S.}~\bibnamefont {Wintz}}, \bibinfo {author} {\bibfnamefont
  {S.}~\bibnamefont {Finizio}}, \bibinfo {author} {\bibfnamefont
  {E.}~\bibnamefont {Josten}}, \bibinfo {author} {\bibfnamefont
  {J.}~\bibnamefont {Raabe}}, \bibinfo {author} {\bibfnamefont
  {M.}~\bibnamefont {Kronseder}}, \bibinfo {author} {\bibfnamefont
  {D.}~\bibnamefont {Bougeard}}, \bibinfo {author} {\bibfnamefont
  {J.}~\bibnamefont {Lindner}},\ and\ \bibinfo {author} {\bibfnamefont {C.~H.}\
  \bibnamefont {Back}},\ }\bibfield  {title} {\bibinfo {title} {Origin and
  manipulation of stable vortex ground states in permalloy nanotubes},\ }\href
  {https://doi.org/10.1021/acs.nanolett.7b05222} {\bibfield  {journal}
  {\bibinfo  {journal} {Nano Letters}\ }\textbf {\bibinfo {volume} {18}},\
  \bibinfo {pages} {2828} (\bibinfo {year} {2018})}\BibitemShut {NoStop}%
\bibitem [{\citenamefont {Vasyukov}\ \emph {et~al.}(2018)\citenamefont
  {Vasyukov}, \citenamefont {Ceccarelli}, \citenamefont {M.~Wyss},
  \citenamefont {Schwarb}, \citenamefont {Mehlin}, \citenamefont {Rossi},
  \citenamefont {Tütüncüoglu}, \citenamefont {Heimbach}, \citenamefont
  {Zamani}, \citenamefont {Kovács}, \citenamefont {i~Morral}, \citenamefont
  {Grundler},\ and\ \citenamefont {Poggio}}]{Vasyukov_NanoLett2018}%
  \BibitemOpen
  \bibfield  {author} {\bibinfo {author} {\bibfnamefont {D.}~\bibnamefont
  {Vasyukov}}, \bibinfo {author} {\bibfnamefont {L.}~\bibnamefont
  {Ceccarelli}}, \bibinfo {author} {\bibfnamefont {B.~G.}\ \bibnamefont
  {M.~Wyss}}, \bibinfo {author} {\bibfnamefont {A.}~\bibnamefont {Schwarb}},
  \bibinfo {author} {\bibfnamefont {A.}~\bibnamefont {Mehlin}}, \bibinfo
  {author} {\bibfnamefont {N.}~\bibnamefont {Rossi}}, \bibinfo {author}
  {\bibfnamefont {G.}~\bibnamefont {Tütüncüoglu}}, \bibinfo {author}
  {\bibfnamefont {F.}~\bibnamefont {Heimbach}}, \bibinfo {author}
  {\bibfnamefont {R.~R.}\ \bibnamefont {Zamani}}, \bibinfo {author}
  {\bibfnamefont {A.}~\bibnamefont {Kovács}}, \bibinfo {author} {\bibfnamefont
  {A.~F.}\ \bibnamefont {i~Morral}}, \bibinfo {author} {\bibfnamefont
  {D.}~\bibnamefont {Grundler}},\ and\ \bibinfo {author} {\bibfnamefont
  {M.}~\bibnamefont {Poggio}},\ }\bibfield  {title} {\bibinfo {title} {Imaging
  stray magnetic field of individual ferromagnetic nanotubes},\ }\href
  {https://doi.org/10.1021/acs.nanolett.7b04386} {\bibfield  {journal}
  {\bibinfo  {journal} {Nano Letters}\ }\textbf {\bibinfo {volume} {18}},\
  \bibinfo {pages} {964} (\bibinfo {year} {2018})}\BibitemShut {NoStop}%
\bibitem [{\citenamefont {Guo}\ \emph {et~al.}(2018)\citenamefont {Guo},
  \citenamefont {Day}, \citenamefont {Chen}, \citenamefont {Tong},
  \citenamefont {Mansikkam{\"a}ki},\ and\ \citenamefont
  {Layfield}}]{Guo_science2018}%
  \BibitemOpen
  \bibfield  {author} {\bibinfo {author} {\bibfnamefont {F.-S.}\ \bibnamefont
  {Guo}}, \bibinfo {author} {\bibfnamefont {B.~M.}\ \bibnamefont {Day}},
  \bibinfo {author} {\bibfnamefont {Y.-C.}\ \bibnamefont {Chen}}, \bibinfo
  {author} {\bibfnamefont {M.-L.}\ \bibnamefont {Tong}}, \bibinfo {author}
  {\bibfnamefont {A.}~\bibnamefont {Mansikkam{\"a}ki}},\ and\ \bibinfo {author}
  {\bibfnamefont {R.~A.}\ \bibnamefont {Layfield}},\ }\bibfield  {title}
  {\bibinfo {title} {Magnetic hysteresis up to 80 kelvin in a dysprosium
  metallocene single-molecule magnet},\ }\href
  {https://doi.org/10.1126/science.aav0652} {\bibfield  {journal} {\bibinfo
  {journal} {Science}\ }\textbf {\bibinfo {volume} {362}},\ \bibinfo {pages}
  {1400} (\bibinfo {year} {2018})}\BibitemShut {NoStop}%
\bibitem [{\citenamefont {Wyss}\ \emph {et~al.}(2017)\citenamefont {Wyss},
  \citenamefont {Mehlin}, \citenamefont {Gross}, \citenamefont {Buchter},
  \citenamefont {Farhan}, \citenamefont {Buzzi}, \citenamefont {Kleibert},
  \citenamefont {T{\"{u}}t{\"{u}}nc{\"{u}}oglu}, \citenamefont {Heimbach},
  \citenamefont {Morral}, \citenamefont {Grundler},\ and\ \citenamefont
  {Poggio}}]{Wyss_prb2017}%
  \BibitemOpen
  \bibfield  {author} {\bibinfo {author} {\bibfnamefont {M.}~\bibnamefont
  {Wyss}}, \bibinfo {author} {\bibfnamefont {A.}~\bibnamefont {Mehlin}},
  \bibinfo {author} {\bibfnamefont {B.}~\bibnamefont {Gross}}, \bibinfo
  {author} {\bibfnamefont {A.}~\bibnamefont {Buchter}}, \bibinfo {author}
  {\bibfnamefont {A.}~\bibnamefont {Farhan}}, \bibinfo {author} {\bibfnamefont
  {M.}~\bibnamefont {Buzzi}}, \bibinfo {author} {\bibfnamefont
  {A.}~\bibnamefont {Kleibert}}, \bibinfo {author} {\bibfnamefont
  {G.}~\bibnamefont {T{\"{u}}t{\"{u}}nc{\"{u}}oglu}}, \bibinfo {author}
  {\bibfnamefont {F.}~\bibnamefont {Heimbach}}, \bibinfo {author}
  {\bibfnamefont {A.~F.}\ \bibnamefont {Morral}}, \bibinfo {author}
  {\bibfnamefont {D.}~\bibnamefont {Grundler}},\ and\ \bibinfo {author}
  {\bibfnamefont {M.}~\bibnamefont {Poggio}},\ }\bibfield  {title} {\bibinfo
  {title} {{Imaging magnetic vortex configurations in ferromagnetic
  nanotubes}},\ }\href {https://doi.org/10.1103/PhysRevB.96.024423} {\bibfield
  {journal} {\bibinfo  {journal} {Phys. Rev. B}\ }\textbf {\bibinfo {volume}
  {024423}},\ \bibinfo {pages} {1} (\bibinfo {year} {2017})}\BibitemShut
  {NoStop}%
\bibitem [{\citenamefont {Mehlin}\ \emph {et~al.}(2018)\citenamefont {Mehlin},
  \citenamefont {Gross}, \citenamefont {Wyss}, \citenamefont {Schefer},
  \citenamefont {T{\"{u}}t{\"{u}}nc{\"{u}}oglu}, \citenamefont {Heimbach},
  \citenamefont {Morral}, \citenamefont {Grundler},\ and\ \citenamefont
  {Poggio}}]{Mehlin_prb2018}%
  \BibitemOpen
  \bibfield  {author} {\bibinfo {author} {\bibfnamefont {A.}~\bibnamefont
  {Mehlin}}, \bibinfo {author} {\bibfnamefont {B.}~\bibnamefont {Gross}},
  \bibinfo {author} {\bibfnamefont {M.}~\bibnamefont {Wyss}}, \bibinfo {author}
  {\bibfnamefont {T.}~\bibnamefont {Schefer}}, \bibinfo {author} {\bibfnamefont
  {G.}~\bibnamefont {T{\"{u}}t{\"{u}}nc{\"{u}}oglu}}, \bibinfo {author}
  {\bibfnamefont {F.}~\bibnamefont {Heimbach}}, \bibinfo {author}
  {\bibfnamefont {A.~F.~i.}\ \bibnamefont {Morral}}, \bibinfo {author}
  {\bibfnamefont {D.}~\bibnamefont {Grundler}},\ and\ \bibinfo {author}
  {\bibfnamefont {M.}~\bibnamefont {Poggio}},\ }\bibfield  {title} {\bibinfo
  {title} {{Observation of end-vortex nucleation in individual ferromagnetic
  nanotubes}},\ }\href {https://doi.org/10.1103/PhysRevB.97.134422} {\bibfield
  {journal} {\bibinfo  {journal} {Phys. Rev. B}\ }\textbf {\bibinfo {volume}
  {97}},\ \bibinfo {pages} {134422} (\bibinfo {year} {2018})}\BibitemShut
  {NoStop}%
\bibitem [{\citenamefont {Fernandez-Roldan}\ \emph
  {et~al.}(2018{\natexlab{b}})\citenamefont {Fernandez-Roldan}, \citenamefont
  {Chrischon}, \citenamefont {Dorneles}, \citenamefont {Chubykalo-Fesenko},
  \citenamefont {Vazquez},\ and\ \citenamefont
  {Bran}}]{Fernandez-Roldan_nanom2018}%
  \BibitemOpen
  \bibfield  {author} {\bibinfo {author} {\bibfnamefont {J.~A.}\ \bibnamefont
  {Fernandez-Roldan}}, \bibinfo {author} {\bibfnamefont {D.}~\bibnamefont
  {Chrischon}}, \bibinfo {author} {\bibfnamefont {L.~S.}\ \bibnamefont
  {Dorneles}}, \bibinfo {author} {\bibfnamefont {O.}~\bibnamefont
  {Chubykalo-Fesenko}}, \bibinfo {author} {\bibfnamefont {M.}~\bibnamefont
  {Vazquez}},\ and\ \bibinfo {author} {\bibfnamefont {C.}~\bibnamefont
  {Bran}},\ }\bibfield  {title} {\bibinfo {title} {{A Comparative Study of
  Magnetic Properties of Large Diameter Co Nanowires and Nanotubes}},\ }\href
  {https://doi.org/10.3390/nano8090692} {\bibfield  {journal} {\bibinfo
  {journal} {Nanomaterials}\ }\textbf {\bibinfo {volume} {8}},\ \bibinfo
  {pages} {692} (\bibinfo {year} {2018}{\natexlab{b}})}\BibitemShut {NoStop}%
\bibitem [{\citenamefont {Landeros}\ \emph {et~al.}(2009)\citenamefont
  {Landeros}, \citenamefont {Suarez}, \citenamefont {Cuchillo},\ and\
  \citenamefont {Vargas}}]{landeros_prb2009}%
  \BibitemOpen
  \bibfield  {author} {\bibinfo {author} {\bibfnamefont {P.}~\bibnamefont
  {Landeros}}, \bibinfo {author} {\bibfnamefont {O.~J.}\ \bibnamefont
  {Suarez}}, \bibinfo {author} {\bibfnamefont {A.}~\bibnamefont {Cuchillo}},\
  and\ \bibinfo {author} {\bibfnamefont {P.}~\bibnamefont {Vargas}},\
  }\bibfield  {title} {\bibinfo {title} {Equilibrium states and vortex domain
  wall nucleation in ferromagnetic nanotubes},\ }\href
  {https://doi.org/10.1103/PhysRevB.79.024404} {\bibfield  {journal} {\bibinfo
  {journal} {Phys. Rev. B}\ }\textbf {\bibinfo {volume} {79}},\ \bibinfo
  {pages} {024404} (\bibinfo {year} {2009})}\BibitemShut {NoStop}%
\bibitem [{\citenamefont {Allende}\ \emph {et~al.}(2008)\citenamefont
  {Allende}, \citenamefont {Escrig}, \citenamefont {Altbir}, \citenamefont
  {Salcedo},\ and\ \citenamefont {Bahiana}}]{Allende_ejpb2008}%
  \BibitemOpen
  \bibfield  {author} {\bibinfo {author} {\bibfnamefont {S.}~\bibnamefont
  {Allende}}, \bibinfo {author} {\bibfnamefont {J.}~\bibnamefont {Escrig}},
  \bibinfo {author} {\bibfnamefont {D.}~\bibnamefont {Altbir}}, \bibinfo
  {author} {\bibfnamefont {E.}~\bibnamefont {Salcedo}},\ and\ \bibinfo {author}
  {\bibfnamefont {M.}~\bibnamefont {Bahiana}},\ }\bibfield  {title} {\bibinfo
  {title} {Angular dependence of the transverse and vortex modes in magnetic
  nanotubes},\ }\href {https://doi.org/10.1140/epjb/e2008-00385-4} {\bibfield
  {journal} {\bibinfo  {journal} {Eur. Phys. J. B}\ }\textbf {\bibinfo {volume}
  {66}},\ \bibinfo {pages} {37} (\bibinfo {year} {2008})}\BibitemShut {NoStop}%
\bibitem [{\citenamefont {Lee}\ \emph {et~al.}(2007)\citenamefont {Lee},
  \citenamefont {Suess}, \citenamefont {Schrefl}, \citenamefont {Oh},\ and\
  \citenamefont {Fidler}}]{Lee_jmmm2007}%
  \BibitemOpen
  \bibfield  {author} {\bibinfo {author} {\bibfnamefont {J.}~\bibnamefont
  {Lee}}, \bibinfo {author} {\bibfnamefont {D.}~\bibnamefont {Suess}}, \bibinfo
  {author} {\bibfnamefont {T.}~\bibnamefont {Schrefl}}, \bibinfo {author}
  {\bibfnamefont {K.~H.}\ \bibnamefont {Oh}},\ and\ \bibinfo {author}
  {\bibfnamefont {J.}~\bibnamefont {Fidler}},\ }\bibfield  {title} {\bibinfo
  {title} {{Magnetic characteristics of ferromagnetic nanotube}},\ }\href
  {https://doi.org/10.1016/j.jmmm.2006.10.1137} {\bibfield  {journal} {\bibinfo
   {journal} {J. Magn. Magn. Mater.}\ }\textbf {\bibinfo {volume} {310}},\
  \bibinfo {pages} {2445} (\bibinfo {year} {2007})}\BibitemShut {NoStop}%
\bibitem [{\citenamefont {Chen}\ \emph {et~al.}(2018)\citenamefont {Chen},
  \citenamefont {Gonzalez},\ and\ \citenamefont {Guslienko}}]{Chen_mater2018}%
  \BibitemOpen
  \bibfield  {author} {\bibinfo {author} {\bibfnamefont {A.~P.}\ \bibnamefont
  {Chen}}, \bibinfo {author} {\bibfnamefont {J.}~\bibnamefont {Gonzalez}},\
  and\ \bibinfo {author} {\bibfnamefont {K.}~\bibnamefont {Guslienko}},\
  }\bibfield  {title} {\bibinfo {title} {{Magnetization reversal modes in short
  nanotubes with chiral vortex domain walls}},\ }\href
  {https://doi.org/10.3390/ma11010101} {\bibfield  {journal} {\bibinfo
  {journal} {Materials}\ }\textbf {\bibinfo {volume} {11}},\ \bibinfo {pages}
  {1} (\bibinfo {year} {2018})}\BibitemShut {NoStop}%
\bibitem [{\citenamefont {Li}\ \emph {et~al.}(2010)\citenamefont {Li},
  \citenamefont {Wang},\ and\ \citenamefont {Li}}]{Li-Ying_pssb2010}%
  \BibitemOpen
  \bibfield  {author} {\bibinfo {author} {\bibfnamefont {Y.}~\bibnamefont
  {Li}}, \bibinfo {author} {\bibfnamefont {T.}~\bibnamefont {Wang}},\ and\
  \bibinfo {author} {\bibfnamefont {Y.}~\bibnamefont {Li}},\ }\bibfield
  {title} {\bibinfo {title} {The influence of dipolar interaction on magnetic
  properties in nanomagnets with different shapes},\ }\href
  {https://doi.org/10.1002/pssb.200945471} {\bibfield  {journal} {\bibinfo
  {journal} {Phys. Status Solidi B}\ }\textbf {\bibinfo {volume} {247}},\
  \bibinfo {pages} {1237} (\bibinfo {year} {2010})}\BibitemShut {NoStop}%
\bibitem [{\citenamefont {Mi}\ \emph {et~al.}(2016)\citenamefont {Mi},
  \citenamefont {Zhai},\ and\ \citenamefont {Hua}}]{Mi_jmmm2016}%
  \BibitemOpen
  \bibfield  {author} {\bibinfo {author} {\bibfnamefont {B.-Z.}\ \bibnamefont
  {Mi}}, \bibinfo {author} {\bibfnamefont {L.-J.}\ \bibnamefont {Zhai}},\ and\
  \bibinfo {author} {\bibfnamefont {L.-L.}\ \bibnamefont {Hua}},\ }\bibfield
  {title} {\bibinfo {title} {Magnon specific heat and free energy of heisenberg
  ferromagnetic single-walled nanotubes: Green's function approach},\ }\href
  {https://doi.org/https://doi.org/10.1016/j.jmmm.2015.09.016} {\bibfield
  {journal} {\bibinfo  {journal} {J. Magn. Magn. Mater.}\ }\textbf {\bibinfo
  {volume} {398}},\ \bibinfo {pages} {160 } (\bibinfo {year}
  {2016})}\BibitemShut {NoStop}%
\bibitem [{\citenamefont {Canko}\ \emph {et~al.}(2014)\citenamefont {Canko},
  \citenamefont {Taşkın}, \citenamefont {Argin},\ and\ \citenamefont
  {Erdinç}}]{Canko_ssc2013}%
  \BibitemOpen
  \bibfield  {author} {\bibinfo {author} {\bibfnamefont {O.}~\bibnamefont
  {Canko}}, \bibinfo {author} {\bibfnamefont {F.}~\bibnamefont {Taşkın}},
  \bibinfo {author} {\bibfnamefont {K.}~\bibnamefont {Argin}},\ and\ \bibinfo
  {author} {\bibfnamefont {A.}~\bibnamefont {Erdinç}},\ }\bibfield  {title}
  {\bibinfo {title} {Hysteresis behavior of blume–capel model on a
  cylindrical ising nanotube},\ }\href
  {https://doi.org/https://doi.org/10.1016/j.ssc.2013.12.020} {\bibfield
  {journal} {\bibinfo  {journal} {Solid State Communications}\ }\textbf
  {\bibinfo {volume} {183}},\ \bibinfo {pages} {35 } (\bibinfo {year}
  {2014})}\BibitemShut {NoStop}%
\bibitem [{\citenamefont {d'Albuquerque~e Castro}\ \emph
  {et~al.}(2002)\citenamefont {d'Albuquerque~e Castro}, \citenamefont {Altbir},
  \citenamefont {Retamal},\ and\ \citenamefont {Vargas}}]{Albuquerque_prl2002}%
  \BibitemOpen
  \bibfield  {author} {\bibinfo {author} {\bibfnamefont {J.}~\bibnamefont
  {d'Albuquerque~e Castro}}, \bibinfo {author} {\bibfnamefont {D.}~\bibnamefont
  {Altbir}}, \bibinfo {author} {\bibfnamefont {J.~C.}\ \bibnamefont
  {Retamal}},\ and\ \bibinfo {author} {\bibfnamefont {P.}~\bibnamefont
  {Vargas}},\ }\bibfield  {title} {\bibinfo {title} {Scaling approach to the
  magnetic phase diagram of nanosized systems},\ }\href
  {https://doi.org/10.1103/PhysRevLett.88.237202} {\bibfield  {journal}
  {\bibinfo  {journal} {Phys. Rev. Lett.}\ }\textbf {\bibinfo {volume} {88}},\
  \bibinfo {pages} {237202} (\bibinfo {year} {2002})}\BibitemShut {NoStop}%
\bibitem [{\citenamefont {Landeros}\ \emph {et~al.}(2005)\citenamefont
  {Landeros}, \citenamefont {Escrig}, \citenamefont {Altbir}, \citenamefont
  {Laroze}, \citenamefont {d'Albuquerque~e Castro},\ and\ \citenamefont
  {Vargas}}]{Landeros_prb2005}%
  \BibitemOpen
  \bibfield  {author} {\bibinfo {author} {\bibfnamefont {P.}~\bibnamefont
  {Landeros}}, \bibinfo {author} {\bibfnamefont {J.}~\bibnamefont {Escrig}},
  \bibinfo {author} {\bibfnamefont {D.}~\bibnamefont {Altbir}}, \bibinfo
  {author} {\bibfnamefont {D.}~\bibnamefont {Laroze}}, \bibinfo {author}
  {\bibfnamefont {J.}~\bibnamefont {d'Albuquerque~e Castro}},\ and\ \bibinfo
  {author} {\bibfnamefont {P.}~\bibnamefont {Vargas}},\ }\bibfield  {title}
  {\bibinfo {title} {Scaling relations for magnetic nanoparticles},\ }\href
  {https://doi.org/10.1103/PhysRevB.71.094435} {\bibfield  {journal} {\bibinfo
  {journal} {Phys. Rev. B}\ }\textbf {\bibinfo {volume} {71}},\ \bibinfo
  {pages} {094435} (\bibinfo {year} {2005})}\BibitemShut {NoStop}%
\bibitem [{\citenamefont {Lehtinen}\ \emph {et~al.}(2004)\citenamefont
  {Lehtinen}, \citenamefont {Foster}, \citenamefont {Ma}, \citenamefont
  {Krasheninnikov},\ and\ \citenamefont {Nieminen}}]{Lehtinen_prl2004}%
  \BibitemOpen
  \bibfield  {author} {\bibinfo {author} {\bibfnamefont {P.~O.}\ \bibnamefont
  {Lehtinen}}, \bibinfo {author} {\bibfnamefont {A.~S.}\ \bibnamefont
  {Foster}}, \bibinfo {author} {\bibfnamefont {Y.}~\bibnamefont {Ma}}, \bibinfo
  {author} {\bibfnamefont {A.~V.}\ \bibnamefont {Krasheninnikov}},\ and\
  \bibinfo {author} {\bibfnamefont {R.~M.}\ \bibnamefont {Nieminen}},\
  }\bibfield  {title} {\bibinfo {title} {Irradiation-induced magnetism in
  graphite: A density functional study},\ }\href
  {https://doi.org/10.1103/PhysRevLett.93.187202} {\bibfield  {journal}
  {\bibinfo  {journal} {Phys. Rev. Lett.}\ }\textbf {\bibinfo {volume} {93}},\
  \bibinfo {pages} {187202} (\bibinfo {year} {2004})}\BibitemShut {NoStop}%
\bibitem [{\citenamefont {Konstantinova}(2008)}]{Konstantinova_jmmm2008}%
  \BibitemOpen
  \bibfield  {author} {\bibinfo {author} {\bibfnamefont {E.}~\bibnamefont
  {Konstantinova}},\ }\bibfield  {title} {\bibinfo {title} {Theoretical
  simulations of magnetic nanotubes using monte carlo method},\ }\href
  {https://doi.org/10.1016/j.jmmm.2008.06.007} {\bibfield  {journal} {\bibinfo
  {journal} {J. Magn. Magn. Mater.}\ }\textbf {\bibinfo {volume} {320}},\
  \bibinfo {pages} {2721 } (\bibinfo {year} {2008})}\BibitemShut {NoStop}%
\bibitem [{\citenamefont {Sotnikov}\ \emph {et~al.}(2017)\citenamefont
  {Sotnikov}, \citenamefont {Mazurenko}, \citenamefont {Katanin}, \citenamefont
  {Mazurenko},\ and\ \citenamefont {Katanin}}]{Sotnikov_prb2017}%
  \BibitemOpen
  \bibfield  {author} {\bibinfo {author} {\bibfnamefont {O.~M.}\ \bibnamefont
  {Sotnikov}}, \bibinfo {author} {\bibfnamefont {V.~V.}\ \bibnamefont
  {Mazurenko}}, \bibinfo {author} {\bibfnamefont {A.~A.}\ \bibnamefont
  {Katanin}}, \bibinfo {author} {\bibfnamefont {V.~V.}\ \bibnamefont
  {Mazurenko}},\ and\ \bibinfo {author} {\bibfnamefont {A.~A.}\ \bibnamefont
  {Katanin}},\ }\bibfield  {title} {\bibinfo {title} {{Monte Carlo study of
  magnetic nanoparticles adsorbed on halloysite Al2Si2O5(OH)4 nanotubes}},\
  }\href {https://doi.org/10.1103/PhysRevB.96.224404} {\bibfield  {journal}
  {\bibinfo  {journal} {Phys. Rev. B}\ }\textbf {\bibinfo {volume} {96}},\
  \bibinfo {pages} {224404} (\bibinfo {year} {2017})}\BibitemShut {NoStop}%
\bibitem [{\citenamefont {Stanković}\ \emph {et~al.}(2019)\citenamefont
  {Stanković}, \citenamefont {Dašić}, \citenamefont {Otálora},\ and\
  \citenamefont {García}}]{Stankovic_nanosc2019}%
  \BibitemOpen
  \bibfield  {author} {\bibinfo {author} {\bibfnamefont {I.}~\bibnamefont
  {Stanković}}, \bibinfo {author} {\bibfnamefont {M.}~\bibnamefont {Dašić}},
  \bibinfo {author} {\bibfnamefont {J.~A.}\ \bibnamefont {Otálora}},\ and\
  \bibinfo {author} {\bibfnamefont {C.}~\bibnamefont {García}},\ }\bibfield
  {title} {\bibinfo {title} {{A platform for nanomagnetism {\textendash}
  assembled ferromagnetic and antiferromagnetic dipolar tubes}},\ }\href
  {https://doi.org/10.1039/C8NR06936K} {\bibfield  {journal} {\bibinfo
  {journal} {Nanoscale}\ }\textbf {\bibinfo {volume} {11}},\ \bibinfo {pages}
  {2521} (\bibinfo {year} {2019})}\BibitemShut {NoStop}%
\bibitem [{\citenamefont {Bogani}\ \emph {et~al.}(2009)\citenamefont {Bogani},
  \citenamefont {Danieli}, \citenamefont {Biavardi}, \citenamefont {Bendiab},
  \citenamefont {Barra}, \citenamefont {Dalcanale}, \citenamefont
  {Wernsdorfer},\ and\ \citenamefont {Cornia}}]{Bogani_angew2009}%
  \BibitemOpen
  \bibfield  {author} {\bibinfo {author} {\bibfnamefont {L.}~\bibnamefont
  {Bogani}}, \bibinfo {author} {\bibfnamefont {C.}~\bibnamefont {Danieli}},
  \bibinfo {author} {\bibfnamefont {E.}~\bibnamefont {Biavardi}}, \bibinfo
  {author} {\bibfnamefont {N.}~\bibnamefont {Bendiab}}, \bibinfo {author}
  {\bibfnamefont {A.-L.}\ \bibnamefont {Barra}}, \bibinfo {author}
  {\bibfnamefont {E.}~\bibnamefont {Dalcanale}}, \bibinfo {author}
  {\bibfnamefont {W.}~\bibnamefont {Wernsdorfer}},\ and\ \bibinfo {author}
  {\bibfnamefont {A.}~\bibnamefont {Cornia}},\ }\bibfield  {title} {\bibinfo
  {title} {Single-molecule-magnet carbon-nanotube hybrids},\ }\href
  {https://doi.org/10.1002/anie.200804967} {\bibfield  {journal} {\bibinfo
  {journal} {Angewandte Chemie - International Edition}\ }\textbf {\bibinfo
  {volume} {48}},\ \bibinfo {pages} {746} (\bibinfo {year} {2009})}\BibitemShut
  {NoStop}%
\bibitem [{\citenamefont {Giusti}\ \emph {et~al.}(2009)\citenamefont {Giusti},
  \citenamefont {Charron}, \citenamefont {Mazerat}, \citenamefont {Compain},
  \citenamefont {Mialane}, \citenamefont {Dolbecq}, \citenamefont
  {Rivi{\`{e}}re}, \citenamefont {Wernsdorfer}, \citenamefont {Biboum},
  \citenamefont {Keita}, \citenamefont {Nadjo}, \citenamefont {Filoramo},
  \citenamefont {Bourgoin},\ and\ \citenamefont {Mallah}}]{Giusti_angew2009}%
  \BibitemOpen
  \bibfield  {author} {\bibinfo {author} {\bibfnamefont {A.}~\bibnamefont
  {Giusti}}, \bibinfo {author} {\bibfnamefont {G.}~\bibnamefont {Charron}},
  \bibinfo {author} {\bibfnamefont {S.}~\bibnamefont {Mazerat}}, \bibinfo
  {author} {\bibfnamefont {J.~D.}\ \bibnamefont {Compain}}, \bibinfo {author}
  {\bibfnamefont {P.}~\bibnamefont {Mialane}}, \bibinfo {author} {\bibfnamefont
  {A.}~\bibnamefont {Dolbecq}}, \bibinfo {author} {\bibfnamefont
  {E.}~\bibnamefont {Rivi{\`{e}}re}}, \bibinfo {author} {\bibfnamefont
  {W.}~\bibnamefont {Wernsdorfer}}, \bibinfo {author} {\bibfnamefont {R.~N.}\
  \bibnamefont {Biboum}}, \bibinfo {author} {\bibfnamefont {B.}~\bibnamefont
  {Keita}}, \bibinfo {author} {\bibfnamefont {L.}~\bibnamefont {Nadjo}},
  \bibinfo {author} {\bibfnamefont {A.}~\bibnamefont {Filoramo}}, \bibinfo
  {author} {\bibfnamefont {J.~P.}\ \bibnamefont {Bourgoin}},\ and\ \bibinfo
  {author} {\bibfnamefont {T.}~\bibnamefont {Mallah}},\ }\bibfield  {title}
  {\bibinfo {title} {{Magnetic bistability of individual single-molecule
  magnets grafted on single-wall carbon nanotubes}},\ }\href
  {https://doi.org/10.1002/anie.200901806} {\bibfield  {journal} {\bibinfo
  {journal} {Angewandte Chemie - International Edition}\ }\textbf {\bibinfo
  {volume} {48}},\ \bibinfo {pages} {4949} (\bibinfo {year}
  {2009})}\BibitemShut {NoStop}%
\bibitem [{\citenamefont {Masotti}\ and\ \citenamefont
  {Caporali}(2013)}]{Masotti_ijms2013}%
  \BibitemOpen
  \bibfield  {author} {\bibinfo {author} {\bibfnamefont {A.}~\bibnamefont
  {Masotti}}\ and\ \bibinfo {author} {\bibfnamefont {A.}~\bibnamefont
  {Caporali}},\ }\bibfield  {title} {\bibinfo {title} {Preparation of magnetic
  carbon nanotubes for biomedical and biotechnological applications},\ }\href
  {https://doi.org/10.3390/ijms141224619} {\bibfield  {journal} {\bibinfo
  {journal} {Int. J. Mol. Sci.}\ }\textbf {\bibinfo {volume} {14}},\ \bibinfo
  {pages} {24619} (\bibinfo {year} {2013})}\BibitemShut {NoStop}%
\bibitem [{\citenamefont {Ji}\ \emph {et~al.}(2019)\citenamefont {Ji},
  \citenamefont {Ma},\ and\ \citenamefont {Tian}}]{Ji_Small2019}%
  \BibitemOpen
  \bibfield  {author} {\bibinfo {author} {\bibfnamefont {M.}~\bibnamefont
  {Ji}}, \bibinfo {author} {\bibfnamefont {N.}~\bibnamefont {Ma}},\ and\
  \bibinfo {author} {\bibfnamefont {Y.}~\bibnamefont {Tian}},\ }\bibfield
  {title} {\bibinfo {title} {3d lattice engineering of nanoparticles by dna
  shells},\ }\href {https://doi.org/10.1002/smll.201805401} {\bibfield
  {journal} {\bibinfo  {journal} {Small}\ }\textbf {\bibinfo {volume} {0}},\
  \bibinfo {pages} {1805401} (\bibinfo {year} {2019})}\BibitemShut {NoStop}%
\bibitem [{\citenamefont {Fittipaldi}\ \emph {et~al.}(2019)\citenamefont
  {Fittipaldi}, \citenamefont {Cini}, \citenamefont {Annino}, \citenamefont
  {Vindigni}, \citenamefont {Caneschi},\ and\ \citenamefont
  {Sessoli}}]{Fittipaldi_NatMater2019}%
  \BibitemOpen
  \bibfield  {author} {\bibinfo {author} {\bibfnamefont {M.}~\bibnamefont
  {Fittipaldi}}, \bibinfo {author} {\bibfnamefont {A.}~\bibnamefont {Cini}},
  \bibinfo {author} {\bibfnamefont {G.}~\bibnamefont {Annino}}, \bibinfo
  {author} {\bibfnamefont {A.}~\bibnamefont {Vindigni}}, \bibinfo {author}
  {\bibfnamefont {A.}~\bibnamefont {Caneschi}},\ and\ \bibinfo {author}
  {\bibfnamefont {R.}~\bibnamefont {Sessoli}},\ }\bibfield  {title} {\bibinfo
  {title} {{Electric field modulation of magnetic exchange in molecular
  helices}},\ }\href {https://doi.org/10.1038/s41563-019-0288-5} {\bibfield
  {journal} {\bibinfo  {journal} {Nature Materials}\ ,\ \bibinfo {pages} {1}}
  (\bibinfo {year} {2019})}\BibitemShut {NoStop}%
\bibitem [{\citenamefont {Jensen}\ and\ \citenamefont
  {Mackintosh}(1991)}]{Jensen1991}%
  \BibitemOpen
  \bibfield  {author} {\bibinfo {author} {\bibfnamefont {J.}~\bibnamefont
  {Jensen}}\ and\ \bibinfo {author} {\bibfnamefont {A.}~\bibnamefont
  {Mackintosh}},\ }\href {http://www.fys.ku.dk/{~}jjensen/Book/Ebook.pdf}
  {\emph {\bibinfo {title} {{Rare earth magnetism : Structures and
  Excitations}}}}\ (\bibinfo  {publisher} {Clarendon Press},\ \bibinfo {year}
  {1991})\BibitemShut {NoStop}%
\bibitem [{\citenamefont {Lehmann}\ \emph {et~al.}(2019)\citenamefont
  {Lehmann}, \citenamefont {Donnelly}, \citenamefont {Derlet}, \citenamefont
  {Heyderman},\ and\ \citenamefont {Fiebig}}]{Lehmann_NatureNano2019}%
  \BibitemOpen
  \bibfield  {author} {\bibinfo {author} {\bibfnamefont {J.}~\bibnamefont
  {Lehmann}}, \bibinfo {author} {\bibfnamefont {C.}~\bibnamefont {Donnelly}},
  \bibinfo {author} {\bibfnamefont {P.~M.}\ \bibnamefont {Derlet}}, \bibinfo
  {author} {\bibfnamefont {L.~J.}\ \bibnamefont {Heyderman}},\ and\ \bibinfo
  {author} {\bibfnamefont {M.}~\bibnamefont {Fiebig}},\ }\bibfield  {title}
  {\bibinfo {title} {{Poling of an artificial magneto-toroidal crystal}},\
  }\href {https://doi.org/10.1038/s41565-018-0321-x} {\bibfield  {journal}
  {\bibinfo  {journal} {Nature Nanotechnology}\ }\textbf {\bibinfo {volume}
  {14}},\ \bibinfo {pages} {141} (\bibinfo {year} {2019})}\BibitemShut
  {NoStop}%
\bibitem [{\citenamefont {Donahue}\ and\ \citenamefont {Porter}(1999)}]{OOMMF}%
  \BibitemOpen
  \bibfield  {author} {\bibinfo {author} {\bibfnamefont {M.}~\bibnamefont
  {Donahue}}\ and\ \bibinfo {author} {\bibfnamefont {D.}~\bibnamefont
  {Porter}},\ }\href@noop {} {\emph {\bibinfo {title} {{OOMMF User’s Guide,
  Version 1.0. Interagency Report NISTIR 6376}}}},\ \bibinfo {type} {Tech.
  Rep.}\ (\bibinfo  {institution} {NIST},\ \bibinfo {year} {1999})\BibitemShut
  {NoStop}%
\bibitem [{\citenamefont {Bran}\ \emph {et~al.}(2013)\citenamefont {Bran},
  \citenamefont {Ivanov}, \citenamefont {Garc{\'{i}}a}, \citenamefont {{Del
  Real}}, \citenamefont {Prida}, \citenamefont {Chubykalo-Fesenko},\ and\
  \citenamefont {Vazquez}}]{Bran_jap2013}%
  \BibitemOpen
  \bibfield  {author} {\bibinfo {author} {\bibfnamefont {C.}~\bibnamefont
  {Bran}}, \bibinfo {author} {\bibfnamefont {Y.~P.}\ \bibnamefont {Ivanov}},
  \bibinfo {author} {\bibfnamefont {J.}~\bibnamefont {Garc{\'{i}}a}}, \bibinfo
  {author} {\bibfnamefont {R.~P.}\ \bibnamefont {{Del Real}}}, \bibinfo
  {author} {\bibfnamefont {V.~M.}\ \bibnamefont {Prida}}, \bibinfo {author}
  {\bibfnamefont {O.}~\bibnamefont {Chubykalo-Fesenko}},\ and\ \bibinfo
  {author} {\bibfnamefont {M.}~\bibnamefont {Vazquez}},\ }\bibfield  {title}
  {\bibinfo {title} {{Tuning the magnetization reversal process of FeCoCu
  nanowire arrays by thermal annealing}},\ }\bibfield  {journal} {\bibinfo
  {journal} {J. App. Phys.}\ }\textbf {\bibinfo {volume} {114}},\ \href
  {https://doi.org/10.1063/1.4816479} {10.1063/1.4816479} (\bibinfo {year}
  {2013})\BibitemShut {NoStop}%
\bibitem [{Sup()}]{Suppl}%
  \BibitemOpen
  \href@noop {} {}\bibinfo {note} {See Supplemental Material at [URL will be
  inserted by publisher] for additional figures.}\BibitemShut {Stop}%
\bibitem [{\citenamefont {Scholz}\ \emph {et~al.}(2003)\citenamefont {Scholz},
  \citenamefont {Guslienko}, \citenamefont {Novosad}, \citenamefont {Suess},
  \citenamefont {Schrefl}, \citenamefont {Chantrell},\ and\ \citenamefont
  {Fidler}}]{Scholz_jmmm2003}%
  \BibitemOpen
  \bibfield  {author} {\bibinfo {author} {\bibfnamefont {W.}~\bibnamefont
  {Scholz}}, \bibinfo {author} {\bibfnamefont {K.}~\bibnamefont {Guslienko}},
  \bibinfo {author} {\bibfnamefont {V.}~\bibnamefont {Novosad}}, \bibinfo
  {author} {\bibfnamefont {D.}~\bibnamefont {Suess}}, \bibinfo {author}
  {\bibfnamefont {T.}~\bibnamefont {Schrefl}}, \bibinfo {author} {\bibfnamefont
  {R.}~\bibnamefont {Chantrell}},\ and\ \bibinfo {author} {\bibfnamefont
  {J.}~\bibnamefont {Fidler}},\ }\bibfield  {title} {\bibinfo {title}
  {Transition from single-domain to vortex state in soft magnetic cylindrical
  nanodots},\ }\href
  {https://doi.org/https://doi.org/10.1016/S0304-8853(03)00466-9} {\bibfield
  {journal} {\bibinfo  {journal} {J. Magn.Magn. Mater.}\ }\textbf {\bibinfo
  {volume} {266}},\ \bibinfo {pages} {155 } (\bibinfo {year}
  {2003})}\BibitemShut {NoStop}%
\end{thebibliography}%

\end{document}